\documentclass[10pt, twocolumn, final]{IEEEtran}%
\usepackage[utf8]{inputenc}

\usepackage[T1]{fontenc}
\usepackage[utf8]{inputenc}
\usepackage{mathtools}

\usepackage{amssymb,mathrsfs}
\usepackage{amsthm}
\usepackage{bm}
\usepackage{scalerel}
\usepackage{nicefrac}
\usepackage{microtype} 
\usepackage[shortlabels]{enumitem}
\usepackage{graphicx}
\usepackage{epstopdf}
\DeclareGraphicsExtensions{.eps,.png,.jpg,.pdf}

\usepackage{url}
\usepackage{colortbl}
\usepackage{booktabs}
\usepackage{multirow}
\usepackage{colortbl,xcolor}
\usepackage{soul}
\usepackage{xparse,xstring}
\usepackage{calc}
\usepackage{etoolbox}

\makeatletter
\@ifpackageloaded{natbib}{
	\relax
}{
	\usepackage{cite}
}
\makeatother

\usepackage{array}
\newcolumntype{L}[1]{>{\raggedright\let\newline\\\arraybackslash\hspace{0pt}}m{#1}}
\newcolumntype{C}[1]{>{\centering\let\newline\\\arraybackslash\hspace{0pt}}m{#1}}
\newcolumntype{R}[1]{>{\raggedleft\let\newline\\\arraybackslash\hspace{0pt}}m{#1}}

\makeatletter
\let\MYcaption\@makecaption
\makeatother
\usepackage[font=footnotesize]{subcaption}
\makeatletter
\let\@makecaption\MYcaption
\makeatother

\usepackage{glossaries}
\makeatletter
\sfcode`\.1006

\let\oldgls\gls
\let\oldglspl\glspl

\newcommand\fussy@ifnextchar[3]{%
	\let\reserved@d=#1%
	\def\reserved@a{#2}%
	\def\reserved@b{#3}%
	\futurelet\@let@token\fussy@ifnch}
\def\fussy@ifnch{%
	\ifx\@let@token\reserved@d
		\let\reserved@c\reserved@a
	\else
		\let\reserved@c\reserved@b
	\fi
	\reserved@c}

\renewcommand{\gls}[1]{%
\oldgls{#1}\fussy@ifnextchar.{\@checkperiod}{\@}}
\renewcommand{\glspl}[1]{%
\oldglspl{#1}\fussy@ifnextchar.{\@checkperiod}{\@}}

\newcommand{\@checkperiod}[1]{%
	\ifnum\sfcode`\.=\spacefactor\else#1\fi
}

\robustify{\gls}
\robustify{\glspl}
\makeatother

\newacronym{wrt}{w.r.t.}{with respect to}
\newacronym{RHS}{R.H.S.}{right-hand side}
\newacronym{LHS}{L.H.S.}{left-hand side}
\newacronym{iid}{i.i.d.}{independent and identically distributed}
\newacronym{SOTA}{SOTA}{state-of-the-art}

\usepackage{float}

\ifx\notloadhyperref\undefined
	\ifx\loadbibentry\undefined
		\usepackage[hidelinks,hypertexnames=false]{hyperref}
	\else
		\usepackage{bibentry}
		\makeatletter\let\saved@bibitem\@bibitem\makeatother
		\usepackage[hidelinks,hypertexnames=false]{hyperref}
		\makeatletter\let\@bibitem\saved@bibitem\makeatother
	\fi
\else
	\ifx\loadbibentry\undefined
		\relax
	\else
		\usepackage{bibentry}
	\fi
\fi

\usepackage[capitalize]{cleveref}
\makeatletter
\def\cref@getref#1#2{%
  \expandafter\let\expandafter#2\csname r@#1@cref\endcsname%
  \expandafter\expandafter\expandafter\def%
    \expandafter\expandafter\expandafter#2%
    \expandafter\expandafter\expandafter{%
      \expandafter\@firstoffive#2}}
\def\cpageref@getref#1#2{%
  \expandafter\let\expandafter#2\csname r@#1@cref\endcsname%
  \expandafter\expandafter\expandafter\def%
    \expandafter\expandafter\expandafter#2%
    \expandafter\expandafter\expandafter{%
      \expandafter\@secondoffive#2}}
      
\AtBeginDocument{%
   \def\label@noarg#1{%
    \cref@old@label{#1}%
    \@bsphack%
    \edef\@tempa{{page}{\the\c@page}}%
    \setcounter{page}{1}%
    \edef\@tempb{\thepage}%
    \expandafter\setcounter\@tempa%
    \cref@constructprefix{page}{\cref@result}%
    \protected@write\@auxout{}%
      {\string\newlabel{#1@cref}{{\cref@currentlabel}%
      {[\@tempb][\arabic{page}][\cref@result]\thepage}{}{}{}}}
    \@esphack}%
  \def\label@optarg[#1]#2{%
    \cref@old@label{#2}%
    \@bsphack%
    \edef\@tempa{{page}{\the\c@page}}%
    \setcounter{page}{1}%
    \edef\@tempb{\thepage}%
    \expandafter\setcounter\@tempa%
    \cref@constructprefix{page}{\cref@result}%
    \protected@edef\cref@currentlabel{%
      \expandafter\cref@override@label@type%
        \cref@currentlabel\@nil{#1}}%
    \protected@write\@auxout{}%
      {\string\newlabel{#2@cref}{{\cref@currentlabel}%
      {[\@tempb][\arabic{page}][\cref@result]\thepage}{}{}{}}}
    \@esphack}%
    }  
\makeatother

\crefname{equation}{}{}
\Crefname{equation}{}{}
\crefname{claim}{claim}{claims}
\crefname{step}{step}{steps}
\crefname{line}{line}{lines}
\crefname{condition}{condition}{conditions}
\crefname{dmath}{}{}
\crefname{dseries}{}{}
\crefname{dgroup}{}{}
\crefname{page}{page}{pages}

\crefname{Problem}{Problem}{Problems}
\crefformat{Problem}{Problem~#2#1#3}
\crefrangeformat{Problem}{Problems~#3#1#4 to~#5#2#6}

\crefname{Theorem}{Theorem}{Theorems}
\crefname{Corollary}{Corollary}{Corollaries}
\crefname{Proposition}{Proposition}{Propositions}
\crefname{Lemma}{Lemma}{Lemmas}
\crefname{Definition}{Definition}{Definitions}
\crefname{Example}{Example}{Examples}
\crefname{Assumption}{Assumption}{Assumptions}
\crefname{Remark}{Remark}{Remarks}
\crefname{Rem}{Remark}{Remarks}
\crefname{remarks}{Remarks}{Remarks}
\crefname{Appendix}{Appendix}{Appendices}
\crefname{Supplement}{Supplement}{Supplements}
\crefname{Exercise}{Exercise}{Exercises}
\crefname{Theorem_A}{Theorem}{Theorems}
\crefname{Corollary_A}{Corollary}{Corollaries}
\crefname{Proposition_A}{Proposition}{Propositions}
\crefname{Lemma_A}{Lemma}{Lemmas}
\crefname{Definition_A}{Definition}{Definitions}

\usepackage{crossreftools}
\ifx\notloadhyperref\undefined
\else
	\relax
\fi

\usepackage{algorithm}
\usepackage{algpseudocode}

\ifx\loadbreqn\undefined
	\relax
\else
	\usepackage{breqn}
\fi

\makeatletter
\def\cleartheorem#1{%
    \expandafter\let\csname#1\endcsname\relax
    \expandafter\let\csname c@#1\endcsname\relax
}
\def\clearthms#1{ \@for\tname:=#1\do{\cleartheorem\tname} }
\makeatother

\ifx\renewtheorem\undefined
	\ifx\useTheoremCounter\undefined
		\newtheorem{Theorem}{Theorem}
		\newtheorem{Corollary}{Corollary}
		\newtheorem{Proposition}{Proposition}
		\newtheorem{Lemma}{Lemma}
	\else
		\newtheorem{Theorem}{Theorem}
		\newtheorem{Corollary}[Theorem]{Corollary}
		
	\fi

	\newtheorem{Definition}{Definition}
	\newtheorem{Example}{Example}

\fi

\theoremstyle{remark}

\theoremstyle{plain}

\newcommand{\qednew}{\nobreak \ifvmode \relax \else
		\ifdim\lastskip<1.5em \hskip-\lastskip
			\hskip1.5em plus0em minus0.5em \fi \nobreak
		\vrule height0.75em width0.5em depth0.25em\fi}



\NewDocumentCommand{\movedownsub}{e{^_}}{%
	\IfNoValueTF{#1}{%
		\IfNoValueF{#2}{^{}}
	}{%
		^{#1}
	}%
	\IfNoValueF{#2}{_{#2}}
}

\let\latexchi\chi
\RenewDocumentCommand{\chi}{}{\latexchi\movedownsub}

\newcommand{\calE}{\mathcal{E}}

\newcommand{\bx}{\mathbf{x}}

\newcommand{\bbR}{\mathbb{R}}

\newcommand{\bbZ}{\mathbb{Z}}

\DeclareSymbolFont{bsfletters}{OT1}{cmss}{bx}{n}
\DeclareSymbolFont{ssfletters}{OT1}{cmss}{m}{n}
\DeclareMathSymbol{\bsfGamma}{0}{bsfletters}{'000}
\DeclareMathSymbol{\ssfGamma}{0}{ssfletters}{'000}
\DeclareMathSymbol{\bsfDelta}{0}{bsfletters}{'001}
\DeclareMathSymbol{\ssfDelta}{0}{ssfletters}{'001}
\DeclareMathSymbol{\bsfTheta}{0}{bsfletters}{'002}
\DeclareMathSymbol{\ssfTheta}{0}{ssfletters}{'002}
\DeclareMathSymbol{\bsfLambda}{0}{bsfletters}{'003}
\DeclareMathSymbol{\ssfLambda}{0}{ssfletters}{'003}
\DeclareMathSymbol{\bsfXi}{0}{bsfletters}{'004}
\DeclareMathSymbol{\ssfXi}{0}{ssfletters}{'004}
\DeclareMathSymbol{\bsfPi}{0}{bsfletters}{'005}
\DeclareMathSymbol{\ssfPi}{0}{ssfletters}{'005}
\DeclareMathSymbol{\bsfSigma}{0}{bsfletters}{'006}
\DeclareMathSymbol{\ssfSigma}{0}{ssfletters}{'006}
\DeclareMathSymbol{\bsfUpsilon}{0}{bsfletters}{'007}
\DeclareMathSymbol{\ssfUpsilon}{0}{ssfletters}{'007}
\DeclareMathSymbol{\bsfPhi}{0}{bsfletters}{'010}
\DeclareMathSymbol{\ssfPhi}{0}{ssfletters}{'010}
\DeclareMathSymbol{\bsfPsi}{0}{bsfletters}{'011}
\DeclareMathSymbol{\ssfPsi}{0}{ssfletters}{'011}
\DeclareMathSymbol{\bsfOmega}{0}{bsfletters}{'012}
\DeclareMathSymbol{\ssfOmega}{0}{ssfletters}{'012}

\makeatletter
\newcommand*\rel@kern[1]{\kern#1\dimexpr\macc@kerna}
\newcommand*\widebar[1]{%
  \begingroup
  \def\mathaccent##1##2{%
    \rel@kern{0.8}%
    \overline{\rel@kern{-0.8}\macc@nucleus\rel@kern{0.2}}%
    \rel@kern{-0.2}%
  }%
  \macc@depth\@ne
  \let\math@bgroup\@empty \let\math@egroup\macc@set@skewchar
  \mathsurround\z@ \frozen@everymath{\mathgroup\macc@group\relax}%
  \macc@set@skewchar\relax
  \let\mathaccentV\macc@nested@a
  \macc@nested@a\relax111{#1}%
  \endgroup
}
\makeatother

\DeclareMathOperator{\spn}{span}

\DeclareMathOperator{\var}{var}

\DeclareMathOperator{\cov}{cov}

\newcommand{\ifbcdot}[1]{\ifblank{#1}{\cdot}{#1}}

\DeclarePairedDelimiterX\abs[1]{\lvert}{\rvert}{\ifbcdot{#1}}
\DeclarePairedDelimiterX\parens[1]{(}{)}{\ifbcdot{#1}}
\DeclarePairedDelimiterX\brk[1]{[}{]}{\ifbcdot{#1}}
\DeclarePairedDelimiterX\braces[1]{\{}{\}}{\ifbcdot{#1}}
\DeclarePairedDelimiterX\angles[1]{\langle}{\rangle}{\ifblank{#1}{\cdot,\cdot}{#1}}
\DeclarePairedDelimiterX\ip[2]{\langle}{\rangle}{\ifbcdot{#1},\ifbcdot{#2}}
\DeclarePairedDelimiterX\norm[1]{\lVert}{\rVert}{\ifbcdot{#1}}
\DeclarePairedDelimiterX\ceil[1]{\lceil}{\rceil}{\ifbcdot{#1}}
\DeclarePairedDelimiterX\floor[1]{\lfloor}{\rfloor}{\ifbcdot{#1}}

\DeclareFontFamily{U}{matha}{\hyphenchar\font45}
\DeclareFontShape{U}{matha}{m}{n}{
      <5> <6> <7> <8> <9> <10> gen * matha
      <10.95> matha10 <12> <14.4> <17.28> <20.74> <24.88> matha12
      }{}
\DeclareSymbolFont{matha}{U}{matha}{m}{n}
\DeclareFontSubstitution{U}{matha}{m}{n}

\DeclareFontFamily{U}{mathx}{\hyphenchar\font45}
\DeclareFontShape{U}{mathx}{m}{n}{
      <5> <6> <7> <8> <9> <10>
      <10.95> <12> <14.4> <17.28> <20.74> <24.88>
      mathx10
      }{}
\DeclareSymbolFont{mathx}{U}{mathx}{m}{n}
\DeclareFontSubstitution{U}{mathx}{m}{n}

\DeclareMathDelimiter{\vvvert}{0}{matha}{"7E}{mathx}{"17}
\DeclarePairedDelimiterX\vertiii[1]{\vvvert}{\vvvert}{\ifbcdot{#1}}

\DeclarePairedDelimiterXPP\trace[1]{\operatorname{Tr}}{(}{)}{}{\ifbcdot{#1}} 
\DeclarePairedDelimiterXPP\col[1]{\operatorname{col}}{\{}{\}}{}{\ifbcdot{#1}} 
\DeclarePairedDelimiterXPP\row[1]{\operatorname{row}}{\{}{\}}{}{\ifbcdot{#1}} 
\DeclarePairedDelimiterXPP\erf[1]{\operatorname{erf}}{(}{)}{}{\ifbcdot{#1}}
\DeclarePairedDelimiterXPP\erfc[1]{\operatorname{erfc}}{(}{)}{}{\ifbcdot{#1}}
\DeclarePairedDelimiterXPP\KLD[2]{D}{(}{)}{}{\ifbcdot{#1}\, \delimsize\|\, \ifbcdot{#2}} 
\DeclarePairedDelimiterXPP\op[2]{\operatorname{#1}}{(}{)}{}{#2} 

\newcommand{\ud}{\,\mathrm{d}} 

\DeclarePairedDelimiterXPP\indicate[1]{{\bf 1}}{\{}{\}}{}{\ifbcdot{#1}}

\NewDocumentCommand\ofrac{s m}{%
	\IfBooleanTF#1%
	{\dfrac{1}{#2}}%
	{\frac{1}{#2}}%
}
\NewDocumentCommand\ddfrac{s m m}{%
	\IfBooleanTF#1%
	{\dfrac{\mathrm{d} {#2}}{\mathrm{d} {#3}}}%
	{\frac{\mathrm{d} {#2}}{\mathrm{d} {#3}}}%
}
\NewDocumentCommand\ppfrac{s m m}{%
	\IfBooleanTF#1%
	{\dfrac{\partial {#2}}{\partial {#3}}}%
	{\frac{\partial {#2}}{\partial {#3}}}%
}

\providecommand\given{}

\DeclarePairedDelimiterX\Set[2]\{\}{%
\renewcommand\given{\SetSymbol[\delimsize]{#1}}
#2
}
\DeclarePairedDelimiterX\Setc[1]\{\}{%
\renewcommand\given{\SetSymbol{:}}
#1
}

\NewDocumentCommand\set{s o m}{%
	\IfBooleanTF#1%
	{\IfValueTF{#2}{\Set*{#2}{#3}}{\Setc*{#3}}}%
	{\IfValueTF{#2}{\Set{#2}{#3}}{\Setc{#3}}}%
}

\NewDocumentCommand{\evalat}{ s O{\big} m e{_^} }{%
\IfBooleanTF{#1}%
{\left. #3 \right|}{#3#2|}%
\IfValueT{#4}{_{#4}}%
\IfValueT{#5}{^{#5}}%
}

\providecommand\given{}
\DeclarePairedDelimiterXPP\cprob[1]{}(){}{
\renewcommand\given{\nonscript\,\delimsize\vert\allowbreak\nonscript\,\mathopen{}}%
#1%
}
\DeclarePairedDelimiterXPP\cexp[1]{}[]{}{
\renewcommand\given{\nonscript\,\delimsize\vert\allowbreak\nonscript\,\mathopen{}}%
#1%
}

\DeclareDocumentCommand \P { s e{_^} d() g } {%
	\mathbb{P}%
	\IfBooleanTF{#1}%
		{
			\IfValueT{#2}{_{#2}}%
			\IfValueT{#3}{^{#3}}%
			\IfValueTF{#5}{\cprob{#4 \given #5}}{\IfValueT{#4}{\cprob{#4}}}%
		}%
		{
			\IfValueT{#2}{_{#2}}%
			\IfValueT{#3}{^{#3}}%
			\IfValueTF{#5}{\cprob*{#4 \given #5}}{\IfValueT{#4}{\cprob*{#4}}}%
		}%
}

\DeclareDocumentCommand \E { s e{_^} o g } {%
	\mathbb{E}%
	\IfBooleanTF{#1}%
		{
			\IfValueT{#2}{_{#2}}%
			\IfValueT{#3}{^{#3}}%
			\IfValueTF{#5}{\cexp{#4 \given #5}}{\IfValueT{#4}{\cexp{#4}}}%
		}%
		{
			\IfValueT{#2}{_{#2}}%
			\IfValueT{#3}{^{#3}}%
			\IfValueTF{#5}{\cexp*{#4 \given #5}}{\IfValueT{#4}{\cexp*{#4}}}%
		}%
}

\DeclareDocumentCommand \Var { s e{_^} d() g } {%
	\var%
	\IfBooleanTF{#1}%
		{
			\IfValueT{#2}{_{#2}}%
			\IfValueT{#3}{^{#3}}%
			\IfValueTF{#5}{\cprob{#4 \given #5}}{\IfValueT{#4}{\cprob{#4}}}%
		}%
		{
			\IfValueT{#2}{_{#2}}%
			\IfValueT{#3}{^{#3}}%
			\IfValueTF{#5}{\cprob*{#4 \given #5}}{\IfValueT{#4}{\cprob*{#4}}}%
		}%
}

\DeclareDocumentCommand \Cov { s e{_^} d() g } {%
	\cov%
	\IfBooleanTF{#1}%
		{
			\IfValueT{#2}{_{#2}}%
			\IfValueT{#3}{^{#3}}%
			\IfValueTF{#5}{\cprob{#4 \given #5}}{\IfValueT{#4}{\cprob{#4}}}%
		}%
		{
			\IfValueT{#2}{_{#2}}%
			\IfValueT{#3}{^{#3}}%
			\IfValueTF{#5}{\cprob*{#4 \given #5}}{\IfValueT{#4}{\cprob*{#4}}}%
		}%
}

\ExplSyntaxOn
\NewDocumentCommand \dist {m o o} {%
\mathrm{#1}\left(%
	\IfValueT{#3}{%
		\tl_if_blank:nTF{ #3 }{\cdot\, \middle|\, }{#3\, \middle|\, }%
	}
	\IfValueT{#2}{#2}%
\right)%
}
\ExplSyntaxOff

\NewDocumentCommand {\cbrace} {t+ D[]{black} D(){\widthof{#5}} m m } {%
	\begingroup%
		\color{#2}
		\IfBooleanTF{#1}{%
			\overbrace{#4}^%
		}{
			\underbrace{#4}_%
		}%
		{\parbox[c]{#3}{\centering\footnotesize{#5}}}%
	\endgroup%
}

\let\oldforall\forall
\renewcommand{\forall}{\oldforall \, }

\let\oldexist\exists
\renewcommand{\exists}{\oldexist \, }

\makeatletter

\newcommand{\rankcolor}[2]{%
	\expandafter\renewcommand\csname #1\endcsname[1]{%
		\ifblank{##1}{%
			{\color{#2} \textbf{#2}}%
		}{%
			\ifmmode
				\textcolor{#2}{\bm{##1}}%
			\else%
				{\color{#2} \textbf{##1}}%
			\fi	
		}%
	}
}

\providecommand{\first}{}

\rankcolor{first}{red}
\rankcolor{second}{blue}
\rankcolor{third}{cyan}
\makeatother

\graphicspath{{./Figures/}{./figures/}}
\DeclareDocumentCommand{\includeCroppedPdf}{ o O{./Figures/} m }{
	\IfFileExists{#2#3-crop.pdf}{}{%
		\immediate\write18{pdfcrop #2#3.pdf #2#3-crop.pdf}}%
	\includegraphics[#1]{#2#3-crop.pdf}
}

\makeatletter
\newcommand*{\addFileDependency}[1]{
  \typeout{(#1)}
  \@addtofilelist{#1}
  \IfFileExists{#1}{}{\typeout{No file #1.}}
}
\makeatother

\definecolor{gray90}{gray}{0.9}
\def\colorlist{red,blue,brown,cyan,darkgray,gray,lightgray,green,lime,magenta,olive,orange,pink,purple,teal,violet,white,yellow}

\makeatletter
\def\startcomment{[}
\ifx\nohighlights\undefined
	\newcommand{\createcolor}[1]{%
			\expandafter\newcommand\csname #1\endcsname[1]{{\color{#1} ##1}}%
	}
	\newcommand{\msout}[1]{\text{\color{green} \st{\ensuremath{#1}}}}
	\newcommand{\del}[1]{{\color{green}\ifmmode \msout{#1}\else\st{#1}\fi}}
\else
	\newcommand{\createcolor}[1]{%
			\expandafter\newcommand\csname #1\endcsname[1]{%
				\noexpandarg%
				\StrChar{##1}{1}[\firstletter]%
				\if\firstletter\startcomment%
					\relax
				\else%
					##1
				\fi
			}%
	}
	\newcommand{\msout}[1]{}
	\newcommand{\del}[1]{}
\fi

\def\@tempa#1,{%
    \ifx\relax#1\relax\else
        \createcolor{#1}%
        \expandafter\@tempa
    \fi
}
\expandafter\@tempa\colorlist,\relax,
\makeatother

\newcommand{\hhide}[1]{}

\ifx\diagnoselabel\undefined
	\relax
\else
	\makeatletter
	\def\@testdef #1#2#3{%
		\def\reserved@a{#3}\expandafter \ifx \csname #1@#2\endcsname
			\reserved@a  \else
			\typeout{^^Jlabel #2 changed:^^J%
				\meaning\reserved@a^^J%
				\expandafter\meaning\csname #1@#2\endcsname^^J}%
			\@tempswatrue \fi}
	\makeatother
\fi

\newacronym{aects}{a.e.\ cts}{almost everywhere continuous}

\usepackage{tikz-cd} 

\newcommand{\numberthis}{\addtocounter{equation}{1}\tag{\theequation}}
\newcommand{\fraks}{\mathfrak{s}}
\NewDocumentCommand{\stret}{ o }{\IfValueT{#1}{^{#1}}^{\fraks}}
\renewcommand{\first}[1]{\ifblank{#1}{\textbf{bold}}{\textbf{#1}}}

\title{Modeling Sparse Graph Sequences and Signals Using Generalized Graphons}
\author{Feng~Ji, Xingchao Jian and Wee~Peng~Tay,~\IEEEmembership{Senior Member,~IEEE} %
\thanks{The authors are with the School of Electrical and Electronic Engineering, Nanyang Technological University, 639798, Singapore (e-mail: jifeng@ntu.edu.sg, xingchao001@e.ntu.edu.sg, wptay@ntu.edu.sg).}%
}

\begin{document}

\maketitle

\begin{abstract}
Graphons are limit objects of sequences of graphs and used to analyze the behavior of large graphs. Recently, graphon signal processing has been developed to study signal processing on large graphs. A major limitation of this approach is that any sparse sequence of graphs inevitably converges to the zero graphon, rendering the resulting signal processing theory trivial and inadequate for sparse graph sequences.
To overcome this limitation, we propose a new signal processing framework that leverages the concept of generalized graphons and introduces the stretched cut distance as a measure to compare these graphons. Our framework focuses on the sampling of graph sequences from generalized graphons and explores the convergence properties of associated operators, spectra, and signals. Our signal processing framework provides a comprehensive approach to analyzing and processing signals on graph sequences, even if they are sparse. Finally, we discuss the practical implications of our theory for real-world large networks through numerical experiments.
\end{abstract}



\section{Introduction}
Modern data analysis often involves complex structures such as graphs. To effectively model and process signals on graphs, graph signal processing (GSP) has developed a range of tools for tasks like sampling, reconstruction, and filtering \cite{Shu13, San13, Tsi15, OrtFroKov:J18, Ji19, LeuMarMou:J23, JiaFenTay:J23, Ji22}. Additionally, graph neural networks (GNNs) have gained significant attention by introducing non-linearity and providing a deep learning architecture for graph-related tasks \cite{Kip16, JiLeeMen23, KanZhaSon23}. These methods leverage the inherent graph structure to achieve high performance and often exhibit desirable computational properties, such as distributed implementation and robustness to perturbations \cite{Seg17, Cec20, ZheZouDon22, SonKanWan22, ZhaKanSon23, KanZhaSon24}.

However, designing signal processing techniques separately for different large graphs can be computationally expensive. For instance, computing the eigendecomposition of a graph shift operator (GSO) for learning a graph filter can be prohibitive when dealing with large graphs. Therefore, it is natural to consider learning graph filters or GNNs on smaller or moderate-sized graphs and then transferring them to larger graphs. The success of such transfer strategies relies on the underlying similarity between the graphs used for training and testing. For example, \cite{Lev21} studies the transferability of GNNs by modeling a graph as a down-sampled version of a topological space, where graph signals are treated as samples of functions on the space. Similarly, \cite{Lev19} investigates the transferability of graph filters in the context of Cayley smoothness space. 

On the other hand, when considering the scenario where the graph size tends to infinity, it becomes natural to study limit objects known as graphons \cite{BorChaLov:J08, Lov:12} in extremal graph theory.

Recently, \cite{Lua21} proposes a signal processing framework based on the graphon theory. This framework provides a way to explain the transferability of graph filters and GNNs through the graphon method \cite{Rui20, Rui20b, Lua21}. A classical graphon is a symmetric function $W: [0,1]\times [0,1] \to [0,1]$. In the context of graphs, the adjacency matrix $A_G$ of a graph $G=(V,E)$ can be intuitively visualized as a step function on $[0,1]^2$ (see illustration in \cite{Gla15} and \cref{defn:lgv} below). The difference between different graphons is quantified by the cut distance \cite{Lov:12}, which is a weaker metric compared to norms such as $\norm{\cdot}_p$. The graphon method can be understood in several ways:
\begin{itemize}
\item Graphons are limit objects of sequences of finite graphs. More precisely, as the number of nodes tends to infinity, many graph sequences converge to a graphon in the cut distance. This has important signal processing implications such as the convergence of graph frequencies. 
\item A graphon may be interpreted as a probabilistic model from where useful finite graphs are sampled. This consideration is used to study the transferability of graph filters on different graphs sampled from the same graphon.
\end{itemize}

A graphon is suitable for modeling the limit of \emph{dense} graph sequences under the cut distance. However, if the graph sequence is \emph{sparse}, then it is no longer appropriate. Formally, for a graph $G=(V,E)$, its edge density is defined as $\calE(G) = |E|/|V|^2$. A sequence of graphs $(G_n)_{n\geq 0}$ is sparse if $\lim_{n\to \infty} \calE(G_n) = 0$. It is well known \cite{Bor18} that \emph{a sparse sequence converges to the zero graphon in the cut distance, rendering the graphon theory unsuitable to deal with sparse graph sequences}. 

Sparse graph sequences are ubiquitous. For example, a sequence $(G_n)_{n\geq 0}$ is sparse if there is a uniform upper bound of the degrees of the vertices in all $G_n$. Another example is when for each $G_n$, only a very small proportion of nodes are densely connected. Many graph sequences that occur in real applications such as social networks, citation networks, and traffic networks are sparse (cf.\ \cref{eg:cas} based on \cite{Bor18}). The classical graphon approach in \cite{Rui20, Rui20b, Lua21} is insufficient to deal with such graph sequences as the limit is the zero graphon.

\begin{figure}[!htb]
\centering
\includegraphics[scale=0.5, trim={0cm 0cm 0 0.0cm},clip]{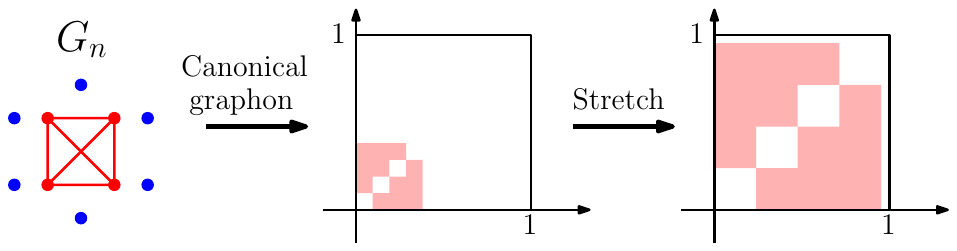}
\caption{Illustration of $G_n$ in \cref{eg:cas}, the canonical graphon (cf.\ \cref{defn:lgv}) and its stretch (cf.\ \cref{eq:stretch}).}
\label{fig:iog}
\end{figure}

\begin{Example} \label{eg:cas}
Consider a sequence of graphs $G_n = (V_n,E_n)$, $n>0$ such that $|V_n|=n$, constructed explicitly as follows. Let $\alpha \in (0,1)$ be a constant. We choose $\floor{n^{\frac{1+\alpha}{2}}}$ vertices $V_n'$ to form a complete subgraph. The remaining $n-|V_n'|$ vertices are isolated vertices (see \cref{fig:iog}). In this case, we have 
\begin{align*}
|E_n| = |V_n'|(|V_n'|-1)/2.
\end{align*}
The edge density $\calE(G_n)$ of $G_n$ is bounded as 
\begin{align*}
\calE(G_n) = |E_n|/n^2 \leq |V_n'|^2/(2n^2). 
\end{align*}
The right-hand-side converges to $0$ since $\alpha<1$. Hence, the graph sequence converges to the zero graphon in the cut distance. The graph $G_n$ is a simplified coarse model for examples such as a social network, where $V_n'$ represents a small number of network celebrities. This example will be analyzed further in \cref{eg:cts} below. We also show examples of real networks in \cref{sec:atl}.
\end{Example}

One solution is to consider generalized graphons on the domain $\mathbb{R}_+^2 = \mathbb{R}_+\times \mathbb{R}_+$, where $\mathbb{R}_+$ is the unbounded set of non-negative numbers \cite{Bor18, Bor19}. Correspondingly, a ``stretch'' procedure turns a sparse sequence of graphs into a denser one that the limit generalized graphon is nonzero. In this way, we can overcome the aforementioned shortcomings. In this paper, we discuss the generalized graphon process and stretched cut distance. Our focus is to present a signal processing framework based on the framework developed in \cite{Bor18, Bor19}. A key objective is to establish a theoretical foundation for the convergence of graphon operators on signals. We remark that though the generalized graphon model is primarily used here to deal with sparse graph sequences, it is versatile and can also be applied to non-sparse sequences. It is shown in \cite{Bor19} that any exchangeable adjacency measure contains a component that is a generalized graphon. For dense graph sequences, the model may lead to different convergence behavior due to the stretching process (e.g., compare \cref{coro:dwm} with \cite[Lemma 4]{Lua21}).

Our main contributions are summarized as follows. 
\begin{itemize}
\item We formulate a general signal processing framework using generalized graphons and the stretched cut distance for sparse graph sequences. 
\item We propose an easily implementable sampling scheme of a sparse graph sequence given a generalized graphon $W$. We show the convergence of the associated canonical graphon (to $W$), and sampled graph signals. 
\item We describe different constructions of filters and discuss their convergence.
\item To verify our theoretical predictions, e.g., spectral convergence, we demonstrate that our proposed model fits better than the classical graphon model on two commonly used benchmark large graph datasets.
\end{itemize}

The rest of the paper is organized as follows. In \cref{sec:gga}, we formally define generalized graphons and introduce the stretched cut distance so that we can compare such objects. In \cref{sec:onc}, we introduce signals on generalized graphons and present the main operator norm convergence result for a sequence of generalized graphons converging in the stretched cut distance. In \cref{sec:sas}, we describe how to obtain converging sequences of graphs given a generalized graphon; while in \cref{sec:sac}, we demonstrate how to obtain converging sequences of graph signals given a general graphon signal. The proposed sample scheme for graphs is simple enough to implement and convenient to analyze in theory. The spectral convergence is discussed in \cref{sec:sai}. We describe what the theory implies for real large networks in \cref{sec:atl} and conclude in \cref{sec:con}. All proofs are in the Appendix.

\emph{Notations}. We use $W$ with appropriate subscripts to denote graphons. Lowercase letters such as $f,g$ denote functions and signals. The Fraktur font $\mathfrak{s}$ is used as the symbol for stretch, and $\square$ is used in notations related to ``cut distance''. $\mathbb{R}_+$ denotes the set of non-negative real numbers and $\mu$ is the Lebesgue measure. The symbol $\norm{\cdot}$ with appropriate subscripts denote different norms. For a finite set $S$, $|S|$ is its size and $1_S$ is the indicator function on the set $S$. The floor function is $\floor{\cdot}$. 

\section{Generalized graphons and the stretched cut distance}\label{sec:gga}

We first review the concept of generalized graphons. The discussion is largely based on \cite{Bor18, Bor19}. A classical graphon \cite{Lov:12} is defined on the compact domain $[0,1]^2 = [0,1]\times [0,1]$. However, in the context of signal processing, it is more practical to define a graphon on a non-compact domain $\Omega^2 = \Omega\times \Omega$. For our purposes, we focus on $\Omega = \mathbb{R}_+$, which is a concrete choice. This allows us to consider a generalized graphon as a non-negative, bounded, and symmetric function on $\mathbb{R}_+^2$ that belongs to $L^1(\mathbb{R}_+^2)$.

\begin{Definition}\label{def:generalized_graphon}
Consider the Lebesgue measure $\mu$ on $\mathbb{R}_+$. A \emph{generalized graphon} $W$ is a non-negative, bounded, and symmetric function on $\mathbb{R}_+^2$. Moreover, we require $W \in L^1(\mathbb{R}_+^2)$. We assume that $\norm{W}_1>0$, unless otherwise stated.
\end{Definition}

As $W$ is assumed to be bounded, it belongs to $L^2(\mathbb{R}_+^2)$ as well \cite[p.126 Remarks (iii)]{Whe77}. 
In many applications, the upper bound of $W$ can be chosen to be $1$, as values of $W(\cdot,\cdot)$ are used as the probability of the existence of edges. 

We denote the $1$-norm and $2$-norm of a generalized graphon $W$ as $\norm{W}_1$ and $\norm{W}_2$ respectively. To handle sparse graph sequences, we introduce the concept of a \emph{stretched graphon} $W^{\mathfrak{s}}$ (cf. \cite{Bor18}), obtained by stretching $W$ with a factor of $\norm{W}_1^{1/2}$. 
The stretched graphon $W^{\mathfrak{s}}$ is defined as
\begin{align}\label{eq:stretch}
W^{\mathfrak{s}}(x,y) = W(\norm{W}_1^{1/2}x,\norm{W}_1^{1/2}y).
\end{align}
This transformation allows us to convert a sparse graphon into a denser one with a fixed $1$-norm of $1$. For convenience, for the rest of the paper, a graphon defined on $[0,1]^2$ is called a \emph{classical graphon}.
In some places, we adopt the following convention if we want to emphasize the stretch factor. 

\emph{Notations}. Given $r>0$ and a function $f$, we use $f^r$ to denote $f$ stretched \gls{wrt} $r$, i.e., $f^r(\bx) = f(r\bx)$. 

An illustration is given in \cref{fig:iog}. The motivation for introducing a stretch is that if $W$ is sparse in the sense that $\norm{W}_1$ is very small, then $W^{\mathfrak{s}}$ is a dense variant whose $1$-norm is always $1$. It is important to note that the addition of two stretched graphons does not yield the same result as stretching the sum of the original graphons, i.e., in general for $W_1, W_2$, we have $(W_1+W_2)^{\mathfrak{s}}\neq W_1^{\mathfrak{s}}+W_2^{\mathfrak{s}}$. 

This exemplifies a subtlety involved in generalizing signal processing methods under the classical graphon framework to the generalized graphon framework. We need to carefully handle stretch factors associated with different graphons and signals, and the results in \cite{Lua21} no longer apply directly. In subsequent sections, we develop additional regularity conditions on graphons and signals. Moreover, on $\mathbb{R}_+$ or $\mathbb{R}_+^2$, we need to (cf.\ \cite{Bor18}) circumvent the non-applicability of certain classical results on compact domains \cite{Lov:12}.


To compare graphons, the cut distance derived from the cut norm is usually used. As we need to consider functions on $\mathbb{R}_+^2$ of the form $W_1-W_2$, we do not require $W$ to be non-negative in the following definition. 

\begin{Definition}\label{def:cut_norm}
The \emph{cut norm} of $W$ is defined as 
\begin{align*}
\norm{W}_{\square} = \sup_{U,V\subset \bbR_+}\abs*{\int_{U\times V}W(x,y)\ud x\ud y},
\end{align*}
where $\sup$ is taken over measurable subsets $U,V$ of $\mathbb{R}_+$. 
\end{Definition}

Key examples of graphons are constructed from finite graphs. 
\begin{Definition} \label{defn:lgv}
Let $G = (V,E)$ be a finite graph of size $n$ with $V = \{v_1\ldots, v_n\}$. The \emph{canonical graphon} $W_G$ associated with $G$ is the step function supported on $[0,1]^2$ as follows. Divide $[0,1]$ into $n$ equal subintervals $(I_l)_{1\leq l\leq n}$ of the same length $1/n$. For $x \in I_i, y\in I_j$, let $W_G(x,y) = 1$ if $(v_i,v_j)\in E$, and $W_G(x,y) = 0$ otherwise. 
In this paper, we extend its domain to $\mathbb{R}_+^2$ by defining $W_G(x,y) = 0$ for $(x,y) \notin [0,1]^2$. 
\end{Definition}
For the canonical graphon associated with $G$, its $1$-norm is 
\begin{align} \label{eq:nw1}
\norm{W_{G}}_1 = 2|E|\cdot \frac{1}{n^2} = 2\calE(G),
\end{align}
as each small square in $[0,1]^2$ has area $1/n^2$ and $W_G \equiv 1$ on $2|E|$ of them. Therefore, $\norm{W_G}_1$ is twice the edge density. 

Apparently, $W_G$ depends on the ordering of the nodes, and in many applications, we do not want to distinguish $W_{G'}$ and $W_G$ if $G'$ is obtained from $G$ by a re-ordering of nodes. To define the cut distance between generalized graphons, we follow \cite{Bor18, Bor19} and allow measure preserving transformations to perturb $W$. More precisely, let $\phi: \mathbb{R}_+ \to \mathbb{R}_+$ be a measure preserving function. Then $W^{\phi}(x,y) = W(\phi(x),\phi(y))$. 

\begin{Definition}
We define $d_{\square}(W_1,W_2) = \norm{W_1-W_2}_{\square}$ and the \emph{cut distance} 
\begin{align*}
\delta_{\square}(W_1,W_2) = \inf_{\phi_1,\phi_2} d_{\square}(W_1^{\phi_1},W_2^{\phi_2}),
\end{align*}
where the infimum is taken over measure preserving functions $\phi_1,\phi_2$. Their stretched versions are defined as
\begin{align*}
& d_{\square,\mathfrak{s}}(W_1,W_2) = d_{\square}(W_1^{\mathfrak{s}},W_2^{\mathfrak{s}}), \text{ and } \\&\delta_{\square,\mathfrak{s}}(W_1,W_2) = \delta_{\square}(W_1^{\mathfrak{s}},W_2^{\mathfrak{s}}).
\end{align*}
The latter is called the \emph{stretched cut distance}. 
\end{Definition}

Notice that converging in $d_{\square}$ implies that in $\delta_{\square}$. Similarly, converging in $d_{\square,\mathfrak{s}}$ implies that in $\delta_{\square,\mathfrak{s}}$. To illustrate, we continue from \cref{eg:cas}.

\begin{Example} \label{eg:cts}
Consider the sparse sequence of graphs $\{G_n=(V_n,E_n),n>0\}$ in \cref{eg:cas}, with the parameter $\alpha\in (0,1)$. Recall that we have a densely connected set of nodes $V_n'$ of size $|V_n'| = \floor{n^{\frac{1+\alpha}{2}}}$. We order $V_n$ by first indexing $V_n'$. Then for the canonical graphon $W_{G_n}$, by (\ref{eq:nw1}), we have 
\begin{align*}
\norm{W_{G_n}}_1 = \frac{|V_n'|(|V_n'|-1)}{n^2} \text{ supported on } \brk*{0, \frac{|V_n'|}{n}}^2.
\end{align*}
For convenience, let
\begin{align*}
\ell_n = \frac{|V_n'|}{\sqrt{|V_n'|(|V_n'|-1)}}.
\end{align*}
The stretched version $W_{G_n}^{\mathfrak{s}}$  of the canonical graphon associated with $G_n$ is $1$ in $[0,\ell_n)^2\backslash D_n$, where $D_n$ consists of $|V_n'|$ diagonal squares each with length $\ell_n/|V_n'|$ (cf.\ \cref{fig:iog}). The stretched graphon $W_{G_n}^{\mathfrak{s}}$ is $0$ elsewhere. As $n\to \infty$, we have 
\begin{align*}
\ell_n \to 1 \text{ and } \mu(D_n) \to 0.
\end{align*}
Therefore, $W_{G_n}^{\mathfrak{s}}$ converges to $I_{[0,1]^2} =I_{[0,1]^2}^{\mathfrak{s}}$ in $\norm{\cdot}_{\square}$, where $I_{[0,1]^2}$ is the indicator function on the unit square. In conclusion, $\{W_{G_n},n>0\}$ converges in the stretched cut distance $\delta_{\square,\mathfrak{s}}$ to a non-zero graphon. On the other hand, $W_{G_n}$ converges to the zero graphon in the cut distance $\delta_{\square}$, as $\norm{W_{G_n}}_1 \to 0$.
\end{Example}

To motivate the notion of graphon signals, recall that in GSP and GNNs, a graph signal $\bx$ on $G=(V,E)$ is a vector in $\mathbb{R}^{|V|}$, where the $i$-th component of $\bx$ is the signal value at the $i$-th node. A graph shift operator $S$ is a linear transformation $\mathbb{R}^{|V|} \to \mathbb{R}^{|V|}$, e.g., $S=A_G$ the adjacency matrix of $G$. It serves as a basic filter on graph signals. Analogously, in the graphon world, objects are no longer discrete, e.g., the finite discrete domain $V$ is replaced either by $[0,1]$ or $\mathbb{R}_+$. Hence accordingly, signals should be replaced by functions, and matrix multiplication should be replaced by integral operations. 

We consider the \emph{graphon signal space} to be $L^2(\mathbb{R_+})$. For any signal $f \in L^2(\mathbb{R_+})$, $W$ defines the function $T_{W^{\mathfrak{s}}}f$ as 
\begin{align} \label{eq:tmf}
T_{W^{\mathfrak{s}}}f(x) = \int_{\mathbb{R}_+} W^{\mathfrak{s}}(x,y)f(y)\ud y.  
\end{align}

We have introduced the main concepts. To end this section, a summary comparison between graphs, classical graphons, and generalized graphons is given in \cref{tab:acb}. 

\begin{table*}[!htb]
\centering
\caption{A comparison between graphs, classical graphons, and generalized graphons.}
\label{tab:acb}
\begin{tabular}{rccc}
\toprule
& Graphs & Classical graphons & Generalized graphons  \\
\midrule
Object & $G$ & $W$ & $W$ (with stretch $W^{\mathfrak{s}}$) \\
\midrule
Domain & $V$ (discrete) & $[0,1]$ & $\mathbb{R}_+$ \\
\midrule
Metric & $\delta_{\square}$ via $W_G$ & $\delta_{\square}$ & $\delta_{\square,\mathfrak{s}}$ \\
\midrule
Signal space & $\mathbb{R}^{|V|}$ & $L^2([0,1])$ & $L^2(\mathbb{R}_+)$ \\
\midrule
Operator (shift) & $A_G$ & $T_W$ & $T_{W^{\mathfrak{s}}}$ \\
\midrule
Relation to graphs & $--$ & Dense limit & Sparse limit \\
\bottomrule
\end{tabular}
\end{table*}

\section{Operator norm convergence} \label{sec:onc}

In this section, we study the convergence of operators on graphon signals. We first introduce the operator norm and then discuss the convergence of operators. To start, we have the following simple observation.

\begin{Lemma} \label{lem:nlf}
For the operator $T_{W^{\mathfrak{s}}}$, we have
\begin{align*}
\norm{T_{W^{\mathfrak{s}}}f}_2 \leq \frac{\norm{W}_2}{\norm{W}_1}\norm{f}_2.
\end{align*}
\end{Lemma}

As a consequence, $W$ induces a \emph{bounded linear operator} $T_{W^{\mathfrak{s}}}: L^2(\mathbb{R}_+) \to L^2(\mathbb{R}_+)$. 
To analyze $T_{W^\mathfrak{s}}$, it is natural to consider the operator norm. 

\begin{Definition}\label{def:op_norm}
Let $\norm{W}_{2,2,\mathfrak{s}}$ denote the operator norm of $T_{W^{\mathfrak{s}}}: L^2(\mathbb{R}_+) \to L^2(\mathbb{R}_+)$. 
Similarly, we define $\norm{W}_{2,2}$ to be the operator norm of $T_W$.
\end{Definition}
From the lemma above, we see  $\norm{W}_{2,2,\mathfrak{s}} \leq \norm{W}_2/\norm{W}_1.$
We have the following regarding operator norm convergence. 

\begin{Theorem} \label{thm:iwi}
If $(W_n)_{n> 0}$ is a sequence of uniformly bounded generalized graphons converging to $W$ in $d_{\square,\mathfrak{s}}$, then they also converge to $W$ in $\norm{\cdot}_{2,2,\mathfrak{s}}$. Moreover, if $(W_n)_{n>0}$ converges to $W$ in $\delta_{\square,\mathfrak{s}}$, then there is a sequence of measure preserving $\phi_n$ such that ${W_n^{\mathfrak{s}}}^{\phi_n} \to W^{\mathfrak{s}}$ in $\norm{\cdot}_{2,2}$. 
\end{Theorem}

In GSP and GNNs, polynomial filters are commonly used. In the context of graphons, we can adopt polynomial filters by considering a polynomial $P$, and obtaining $P(T_{W^{\mathfrak{s}}})$ by plugging $T_{W^{\mathfrak{s}}}$ into $P$, where the (operator) composition serves as the multiplication operation.

\begin{Corollary} \label{coro:lpb}
Let $P$ be a polynomial. If $(W_n)_{n> 0}$ is a sequence of uniformly bounded generalized graphons converging to $W$ in $d_{\square,\mathfrak{s}}$, then $\big(P(T_{W_n^{\mathfrak{s}}})\big)_{n>0}$ converges to $P(T_{W^{\mathfrak{s}}})$ in the operator norm $\norm{\cdot}_{2,2}$ as operators $L^2(\mathbb{R}_+) \to L^2(\mathbb{R}_+)$. 
\end{Corollary}

In practice, we also need to deal with the situation when the signal $f$ is approximated by a sequence $(f_n)_{n> 0}$.

\begin{Corollary} \label{coro:swi}
Suppose $(W_n)_{n> 0}$ is a sequence of uniformly bounded generalized graphons converging to $W$ in $d_{\square,\mathfrak{s}}$. If $(f_n)_{n> 0}$ is a sequence in $L^2(\mathbb{R}_+)$ that converges to $f \in L^2(\mathbb{R}_+)$, then $(T_{W_n^{\mathfrak{s}}}f_n)_{n > 0}$ converges to $T_{W^{\mathfrak{s}}}f$ in $L^2(\mathbb{R}_+)$.  
\end{Corollary}

We aim to obtain samples of finite graphs with associated graph signals from a given generalized graphon and a signal. \Cref{coro:swi} suggests two directions: based on convergence in $d_{\square,\mathfrak{s}}$ (or $\delta_{\square,\mathfrak{s}}$) of generalized graphons and the $2$-norm convergence of signals, respectively. We explore these directions in detail in the following two sections.

\section{Sampling and stretched cut distance convergence} \label{sec:sas}

In this section, we present a sampling procedure that generates a (double) graph sequence converging to a generalized graphon $W$ in the stretched cut distance. 

Suppose that $W: \mathbb{R}_+^2 \to [0,1]$ is a generalized graphon. Let $(t_m)_{m>0}$ be a strictly increasing sequence of positive real numbers such that $t_m \to \infty$ as $m \to \infty$. For each $m$, we sample a sequence of graphs $(G_{m,n})_{n>0}$ as in \cref{alg:wur}.
\begin{algorithm}
\caption{Sampling a double sequence from $W$.}\label{alg:wur}
\begin{itemize}
\item We uniformly randomly sample $n$ independent points $X_{m,n}=\{x_{m,1},\ldots,x_{m,n}\}$ in $[0,t_m]$. 
\item The graph $G_{m,n}$ has $n$ vertices $V_{m,n}= \{v_{m,1}, \ldots, v_{m,n}\}$ and there is an edge $(v_{m,i},v_{m,j})$ with probability $W(x_{m,i},x_{m,j})$. 
\item Let $W_{m,n}$ be the canonical graphon on $[0,1]^2$ associated with $G_{m,n}$.
\end{itemize}
\end{algorithm}

We have the following convergence result regarding the sampled double sequence of graphs. 

\begin{Theorem} \label{thm:swi}
Suppose $W$ is \gls{aects} on $\mathbb{R}_+^2$. Let $W_{m,n}$ be the canonical graphon associated with $G_{m,n}$ in \cref{alg:wur}.
Then with probability one, we have 
\begin{align*}
\lim_{m \to\infty}\lim_{n \to\infty} W_{m,n} = W
\end{align*}
in the stretched cut distance $\delta_{\square,\mathfrak{s}}$. 
\end{Theorem}

The following terms are used in the proof in Appendix~\ref{app:pot} and discussions in \cref{sec:atl}:
\begin{align} \label{eq:www}
\begin{aligned} 
W_m = W1_{[0,t_m]^2}, \text{ and } W_m'(x,y) = W_m(t_mx,t_my),
\end{aligned}
\end{align}
where $1_{[0,t_m]^2}$ is the indicator function of $[0,t_m]^2$.

It is worth pointing out that the sampling procedure in \cref{alg:wur} does not generate points $X_{m,n}$ in order from $[0,t_m]$. Therefore, \cref{thm:swi} concludes only convergence in $\delta_{\square,\mathfrak{s}}$. In practice, to apply \cref{thm:swi}, one may first order the sampled points in $X_{m,n}$ according to their values. 



In \cite{Bor18}, a Poisson point process on $\mathbb{R}_+^2$ is used to sample a sequence of graphs converging to $W$ in the stretched cut distance with probability one. Notice that on each measurable set $U$ with finite measure, the Poisson point process uniformly samples $k$ points with $k$ following the Poisson distribution with parameter $\mu(U)$, which is a feature common to our scheme. 

\cref{alg:wur} is simple to implement. Hence, our approach is straightforward to apply in practice. It is also convenient for us to develop a signal processing theory in the next section. Moreover, if $W$ is compactly supported, then \cite{Bor18} samples a dense sequence of graphs with probability one. However, we have the following. 

\begin{Corollary} \label{coro:ted}
With probability one, edge densities of the double sequence $(G_{m,n})_{m>0,n>0}$ in \cref{alg:wur} converge to~$0$.
\end{Corollary}

This result implies that we can always find sparse subsequences $G_{m,\phi(m)}$ of graphs that converge to $W$ in the stretched cut distance. In view of the proof in Appendix~\ref{app:pot}, this can be done in the following simple but generic way: 
\begin{algorithm}
\caption{Sampling a sparse sequence from $(W_m)_{m>0}$.}\label{alg:fem}

\begin{itemize}
\item For each $m$, choose $\phi(m)$ sufficiently large such that the edge density of $G_{m,\phi(m)}$ differs from $\norm{W_m}_1/t_m^2$ by at most $1/m$.
\item Enlarge $\phi(m)$ such that the stretched cut distance between $W_{m,\phi(m)}$ and $W_m$ is bounded by $1/m$.
\end{itemize}
\end{algorithm}

The additional information of the sequence being ``sparse'' allows us to decide the most appropriate model (involving choosing scaling factors) for signal processing tasks such as spectral analysis and filtering (cf.\ \cref{sec:atl}).

We shall discuss another consequence of \cref{thm:swi} regarding spectral convergence in \cref{sec:sai}. 

\section{Sampling and convergence of signals} \label{sec:sac}

In this section, we discuss the convergence of graph signals. We consider bounded and \gls{aects} graphon signals $f$ defined on $\mathbb{R}_+$. Let $b_f$ be a bound for $f$, and note that $f$ belongs to the space $L^2(\mathbb{R}_+)$. 

Following the notations of \cref{sec:sas}, for $m,n>0$, we construct a sequence of graph signals $g_{m,n}$ on $G_{m,n}$ as follows. Recall that in \cref{alg:wur}, we uniformly randomly sample points $X_{m,n} = \{x_{m,1},\ldots, x_{m,n}\}$ from $[0,t_m]$. Assume they are ordered in increasing order. We define $g_{m,n}(v_{m,i}) = f(x_{m,i})$. Let $f_{m,n}$ be the associated \emph{canonical step function} on $[0,1]$. Specifically, we subdivide $[0,1]$ into $n$ equal intervals $I_1,\ldots, I_n$ of size $1/n$ each. For $x\in I_i$, we define:
\begin{align} \label{eq:fmg}
f_{m,n}(x) = g_{m,n}(v_{m,i})=f(x_{m,i})
\end{align}

To state the main result, let $f_{m,n}^{\mathfrak{s}}$ be the stretched signal of $f_{m,n}$ \gls{wrt} the stretch factor $\norm{W_{m,n}}_1^{1/2}$, i.e., $f_{m,n}^{\mathfrak{s}}(x)=f_{m,n}(\norm{W_{m,n}}_1^{1/2}x)$. We use $f^{\mathfrak{s}}$ to denote $f$ stretched \gls{wrt} the stretch factor $\norm{W}_1^{1/2}$.

\begin{Theorem} \label{thm:sfi}
Suppose $f$ is \gls{aects}. Then with probability one, $\lim_{m \to\infty}\lim_{n \to\infty} f_{m,n}^{\mathfrak{s}} = f^{\mathfrak{s}}$ in $1$-norm.
\end{Theorem}

As all the functions involved are bounded, convergence in $1$-norm in the conclusion also implies convergence in $2$-norm. Therefore, \cref{coro:swi} can be applied to the sampled sequence of signals.

\section{Spectral and filter convergence} \label{sec:sai}

In this section, we discuss the spectral convergence of the sampled double sequence of graphs in \cref{sec:sas}. We follow the symbols and notations of \cref{sec:sas}. In particular, graphs $(G_{m,n} = \big(V_{m,n},E_{m,n})\big)_{m>0,n>0}$ sampled from a generalized graphon $W$ that is \gls{aects} form a double sequence, whose associated canonical graphons (on $[0,1]^2$) $(W_{m,n})_{m>0,n>0}$ converge to $W$ in the stretched cut distance with probability one (cf.\ \cref{thm:swi}).

Recall $T_{W^{\mathfrak{s}}}$ defined by \cref{eq:tmf}. Since we assume that $W$ is symmetric and it belongs to $L^2(\mathbb{R}_+^2)$, so is $W^{\mathfrak{s}}$. Hence, $T_{W^{\mathfrak{s}}}$ is a self-adjoint Hilbert-Schmidt (HS) operator. By the spectral theorem, it admits an orthogonal eigenbasis and the only accumulation point of the set of eigenvalues is zero \cite{Lax02}. Moreover, each non-zero eigenvalue has a finite multiplicity. 

We denote the eigenvalues and orthonormal eigenvectors of $T_W$ as $\Lambda_W = \set{\lambda_t: t\neq 0}$ and $\Phi_W = \set{\varphi_t(x): t\neq 0}$.
The eigenvalues $\Lambda_W = \set{\lambda_t: t\neq 0}$ are ordered such that 
\begin{align*}
\lambda_1(W)\geq\lambda_2(W)\geq\dots0, \text{ and } \lambda_{-1}(W)\leq\lambda_{-2}(W)\leq\dots 0.
\end{align*}
It can be shown by \cite[Theorem 4.2.16]{Dav:07} that $W$ can be decomposed as 
\begin{align}\label{eq:graphex_decomp_funcs}
W(x,x') = \sum_{t\in\bbZ\backslash\{0\}} \lambda_t\varphi_t(x)\varphi_t(x').
\end{align}

Consider the sampled graph $G_{m,n}$ from \cref{alg:wur}. For each pair $m,n$, consider the eigenvalues of the adjacency matrix $A_{m,n}$ of $G_{m,n}$. More specifically, we label the eigenvalues of the symmetric matrix $A_{m,n}$ by 
\begin{align*}
\lambda_{-1,m,n} \leq \lambda_{-2,m,n} \leq \ldots \leq 0 \leq \ldots \leq \lambda_{2,m,n} \leq \lambda_{1,m,n}. 
\end{align*}
We have the following spectral convergence result (cf.\ \cite{JianJiTay:J23b}).

\begin{Corollary}\label{coro:dwm}
Under the same assumptions as \cref{thm:swi}, with probability one, for each $t$, the following holds:
\begin{align}\label{eq:g_freq_conv}
\lim_{m \to\infty}\lim_{n \to\infty} \frac{\lambda_{t,m,n}}{\sqrt{2|E_{m,n}|}} = \frac{\lambda_t}{\sqrt{\norm{W}_1}}. 
\end{align} 
\end{Corollary}
We end this section by generalizing \cref{coro:lpb}. For a general function $h$ such that $h(0)=0$, we define an operator $h(T_{W^{\mathfrak{s}}})$ by using the eigenvectors $\Phi_{W^{\mathfrak{s}}} = \{\overline{\varphi_t}: {t\neq 0}\}$ and eigenvalues $\Lambda_{W^{\mathfrak{s}}} = \{\overline{\lambda_t}:t\neq 0\}$ of $T_{W^{\mathfrak{s}}}$ as
\begin{align}
h(T_{W^{\mathfrak{s}}}) f= \sum_{t\in \mathbb{Z}\backslash \{0\}} h(\overline{\lambda_t})\langle f, \overline{\varphi_t} \rangle \overline{\varphi_t}.
\end{align}
Notice $\overline{\varphi_t}(x) = \varphi_t(\sqrt{\norm{W}_1}x)$ 
and $\overline{\lambda_t}=\lambda_t/\sqrt{\norm{W}_1}$. In theory, the condition $h(0)=0$ is general enough. Since if $h(0)\neq 0$, we can simply replace $h$ with $h - h(0)$.

\begin{Theorem} \label{coro:ihi}
Suppose $h$ is continuous on the closure $\overline{U}$ of a bounded open interval $U$ containing $\Lambda_{W^{\mathfrak{s}}}$. Then with probability one, the following holds (in the operator norm):
\begin{align} \label{eq:lml}
\lim_{m\to\infty}\lim_{n\to\infty} h(T_{W_{m,n}^{\mathfrak{s}}}) = h(T_{W^{\mathfrak{s}}}).
\end{align} 
\end{Theorem}

\Cref{coro:ihi} is a counterpart to \cite[Theorem 4]{Lua21} (on the convergence of Lipshitz continuous filters) for the double sequence $(W_{m,n})_{m>0,n>0}$. Our proof does not rely on eigenvector convergence, which requires additional assumptions on $W$. In practice, it can be challenging and costly to obtain analytic expressions for the eigenvectors of the stretched graphons and sampled large graphs. Therefore ideally, performing explicit eigendecompositions should be avoided if possible. Our proof relies on the Stone-Weierstrass theorem \cite{Rud76}, which uses polynomial filters to approximate $h$. This suggests that polynomial filters, which are easy to handle computationally, are sufficient to approximate more general filter convergence in conjunction with sampled graphons. 

\section{Large networks in practice} \label{sec:atl}

Now we discuss what the theory implies for large networks in practice. Let $G=(V,E)$ be a very large graph of size $n_0$ such that it is costly to perform signal processing or machine learning tasks directly on $G$. We fix a generalized graphon $\overline{W}_G$ associated with $G$. This can be either the canonical graphon $W_G$ or the normalized graphon $\widetilde{W}_G$ associated with the normalized adjacency matrix of $G$. More precisely, the interval $[0,1]$ is partitioned into $n_0$ subintervals of equal size $1/n_0$. For $j\leq n_0$, the $j$-th interval $I_j$ corresponds to the $j$-th node of $G$. Recall for the canonical graphon, $W_G(x,y) = 1$ if $(x,y) \in I_i\times I_j$ for $(v_i,v_j)\in E$, and $0$ otherwise. For the normalized graphon, $\widetilde{W}_G(x,y) = d_i^{-1}d_j^{-1}$ if $(x,y) \in I_i\times I_j$ for $(v_i,v_j)\in E$, where $d_i$ (resp.\ $d_j$) is the degree of $v_i$ (resp.\ $v_j$). In either case, $\overline{W}_G$ is supported on $[0,1]^2$.\footnote{It is sometimes possible to find a smooth approximation of $W$ and extrapolate it to the entire $\mathbb{R}_+\times \mathbb{R}_+$.} 

Recall that we have described how to construct a convergent sequence of sparse graphs $G_{m,\phi(m)}$ at the end of \cref{sec:sas}. We interpret $G=G_{m_0,\phi(m_0)}$ for some large $m_0$. We are interested in finding $G_{m,\phi(m)}$ for $m \ll m_0$, which is computationally more tractable. We assume that $\overline{W}_G$ is a good approximation of $W_{m_0}'$ (cf. (\ref{eq:www})) and $\phi(m) \ll \phi(m_0)$ is known. To obtain $G_m$, directly translating \cref{alg:wur} gives the following procedure:
\begin{itemize}
\item Choose a small hyperparameter $\epsilon_m \geq 0$ (which can be $0$).
\item Construct $\overline{W}_G1_{[0,1-\epsilon_m]^2}$ rescaled to $[0,1]^2$ as an approximation of $W_m'$.
\item Sample a graph $G_{m,\phi(m)}$ with $\phi(m)$  nodes from the rescaled $\overline{W}_G1_{[0,1-\epsilon_m]^2}$. In the experiments below, the subgraph edge densities decay very fast, and $\phi(m)$ is set to be a constant multiple of $m$.   
\end{itemize}
Suppose we simplify by letting $\epsilon_m=0$. The generalized graphon $\overline{W}_G1_{[0,1-\epsilon_m]^2}=\overline{W}_G$ determines an edge for $x \in I_i,y \in I_j$ if $(v_i,v_j)$ is an edge of $G$. Moreover, each subinterval $I_i$ of $[0,1]$ corresponds to a node $v_i$ of $G$. Therefore, the above sampling procedure from $[0,1]$ is the same as uniformly random sampling of nodes from $G$. In the studies below, we may slightly modify this, depending on the specific task, to obtain growing graph sequences converging to $G$.     

The signal on $G_m = G_{m,\phi(m)}$ can be obtained accordingly if we are given a signal on $G_{m_0,\phi(m_0)}=G$, namely, if $G_m$ contains $v\in G$, then the signal on $v$ is inherited from $G$ to $G_m$. \cref{coro:swi} can be interpreted as a transferability result. In the following subsections, we consider some real large networks. 

\subsection{Ogbn-arXiv: spectral convergence}

\begin{table*}[!htb]
\centering
\caption{MSE of linear fit: tail parts. Each entry is obtained by averaging on 20 realizations of $\{G_n\}_{n>0}$. The best performance is in \first{}.} 
\label{tab:MSE}
\begin{tabular}{ p{4cm}p{0.8cm}p{0.8cm}p{0.8cm}p{0.8cm}p{1.3cm} p{0.8cm}p{0.8cm}p{0.8cm}p{0.8cm}p{0.8cm}}
\toprule
& \multicolumn{7}{c}{MSE of linear fit using $G_n$ with $|V_n|\geq 6600$} \\
\midrule
& $\lambda_{1,n}$ & $\lambda_{2,n}$ & $\lambda_{3,n}$ & $\lambda_{4,n}$ & $\lambda_{5,n}$ & $\lambda_{-1,n}$ & $\lambda_{-2,n}$ & $\lambda_{-3,n}$ & $\lambda_{-4,n}$ & $\lambda_{-5,n}$ \\
\midrule
Generalized graphon (ours) & \first{1.06} & \first{0.42} & \first{0.33} & \first{0.22} & \first{0.14} & \first{1.05} & \first{0.46} & \first{0.36} & \first{0.27}& \first{0.17} \\
Classical graphon \cite{Lua21} &  4.38& 1.87 & 1.33& 0.99& 0.82 & 4.09 & 1.84 & 1.22& 0.93& 0.77\\
Graphing \cite{RodGamBar:J22} &  2.18& 1.20 & 1.12& 0.95& 0.73 &  1.69& 1.01 & 0.96& 0.91& 0.60\\ 
\bottomrule
\end{tabular}

\begin{tabular}{ p{4cm}p{0.8cm}p{0.8cm}p{0.8cm}p{0.8cm}p{1.3cm} p{0.8cm}p{0.8cm}p{0.8cm}p{0.8cm}p{0.8cm}}
\toprule
& \multicolumn{7}{c}{MSE of linear fit using $G_n$ with $|V_n| \geq 8000$} \\
\midrule
& $\lambda_{1,n}$ & $\lambda_{2,n}$ & $\lambda_{3,n}$ & $\lambda_{4,n}$ & $\lambda_{5,n}$ & $\lambda_{-1,n}$ & $\lambda_{-2,n}$ & $\lambda_{-3,n}$ & $\lambda_{-4,n}$ & $\lambda_{-5,n}$ \\
\midrule
Generalized graphon (ours) & \first{0.63} & \first{0.12} & \first{0.20} & \first{0.07} & \first{0.06}  & \first{0.58} & \first{0.12} & \first{0.35} & \first{0.05}& \first{0.05}  \\
Classical graphon \cite{Lua21} & 1.40 & 0.48 & 0.43 & 0.27 & 0.24 & 1.26 & 0.46 & 0.53 & 0.24 & 0.20 \\
Graphing \cite{RodGamBar:J22} & 0.93 & 0.27 & 0.40 & 0.23 & 0.18 & 0.77 & 0.21 & 0.56& 0.13 & 0.15\\ 

\bottomrule
\end{tabular}

\begin{tabular}{ p{4cm}p{0.8cm}p{0.8cm}p{0.8cm}p{0.8cm}p{1.3cm} p{0.8cm}p{0.8cm}p{0.8cm}p{0.8cm}p{0.8cm}}
\toprule
& \multicolumn{7}{c}{MSE of linear fit using $G_n$ with $|V_n|\geq 8600$} \\
\midrule
& $\lambda_{1,n}$ & $\lambda_{2,n}$ & $\lambda_{3,n}$ & $\lambda_{4,n}$ & $\lambda_{5,n}$ & $\lambda_{-1,n}$ & $\lambda_{-2,n}$ & $\lambda_{-3,n}$ & $\lambda_{-4,n}$ & $\lambda_{-5,n}$  \\
\midrule
Generalized graphon (ours) & \first{0.69} & \first{0.06} & \first{0.22} & \first{0.04} & \first{0.03} & \first{0.62} & \first{0.05} & \first{0.43} & \first{0.02}& \first{0.02}  \\
Classical graphon \cite{Lua21} &  0.95& 0.20 & 0.30& 0.11& 0.09 & 0.86 & 0.19 & 0.49& 0.09& 0.07 \\
Graphing \cite{RodGamBar:J22} &  0.91& 0.12 & 0.36& 0.12& 0.09 &  0.77& 0.09 & 0.59& 0.08& 0.08 \\ 
\bottomrule
\end{tabular}
\end{table*}


\begin{figure}[!htb]
\centering
\includegraphics[scale=0.3, trim={0cm 0cm 0 0cm},clip]{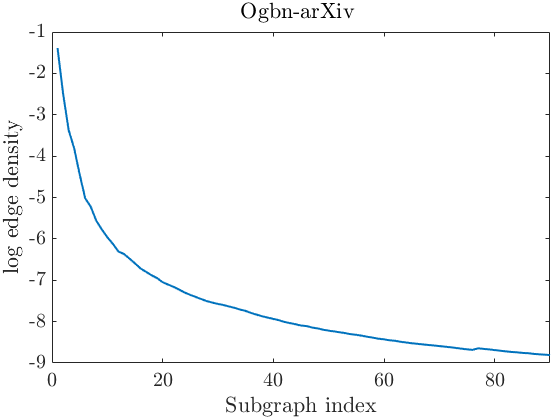}
\caption{The edge densities ($\log$) of (growing) subgraphs of the Ogbn-arXiv graph. 
We see that the edge density decreases as the sampled subgraph grows, suggesting that the graph sequence is sparse.} 
\label{fig:edl}
\end{figure}

\begin{figure}
    \centering
    \includegraphics[scale=0.3, trim={0cm 0cm 0 0cm},clip]{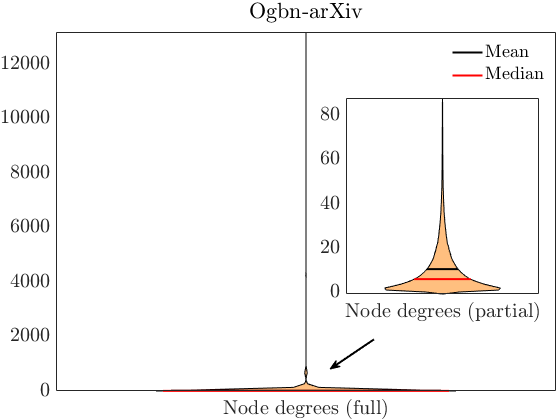}
    \caption{For the node degree distribution of the Ogbn-arXiv graph, we show both the full distribution and the partial distribution where $1.5\%$ nodes with the largest degrees are removed. We see a $\log$-tailed pattern from the plot.}
    \label{fig:tnd}
\end{figure}

In this subsection, we corroborate the spectral convergence result in \cref{sec:sai} on the citation network Ogbn-arxiv.\footnote{\url{https://ogb.stanford.edu/docs/nodeprop/}} Both \cite{Lua21} and \cite{RodGamBar:J22} have spectral convergence results based on their respective graph convergence models (as we discuss below). By studying spectral convergence, we evaluate which model is most appropriate for Ogbn-arxiv. 

The graph, denoted by $G = (V, E)$, has size $|V|\approx 170,000$. We generate a growing graph sequence $G_n\subset G, n\geq1$ as: 
\begin{itemize}
\item We start with an empty graph $\widetilde{G}_0=(\widetilde{V}_0,\widetilde{E}_0)$ with no nodes or edges.
\item For each $n\geq0$, given $\widetilde{G}_n = (\widetilde{V}_n,\widetilde{E}_n)$, we randomly select $200$ nodes from $V\backslash\widetilde{V}_n$ without replacement, and add them into $\widetilde{V}_n$ to obtain $\widetilde{V}_{n+1}$. By letting $\widetilde{E}_{n+1} = E\bigcap(\widetilde{V}_{n+1}\times\widetilde{V}_{n+1})$ we obtain $\widetilde{G}_{n+1}= (\widetilde{V}_{n+1},\widetilde{E}_{n+1})$. We iterate this step for $90$ times to obtain $(\widetilde{G}_n)_{n=1}^{90}$.
\item We remove all isolated vertices in every $\widetilde{G}_n$, and obtain the sequence $\big(G_n=(V_n,E_n)\big)_{1\leq n\leq 90}$ with $|V_{90}|\approx 9500$. 
\end{itemize} 

We consider the finite sequence $(G_n)_{1\leq n\leq 90}$ as sampled from the same graphon that gives $G$. If it converges to a \emph{non-zero} generalized graphon $W$ in the stretched cut distance, then according to \cref{coro:dwm}, the eigenvalues $\set{\lambda_{t,n}\given 1\leq n\leq 90}$ satisfy
\begin{align*}
  \lambda_{t,n} = k_t\sqrt{\abs{E_n}} + o(\sqrt{\abs{E_n}}),    
\end{align*}
as $n$ increases. Here, $o(\cdot)$ is the little-o notation. If $k_t = \lambda_t/\sqrt{\norm{W}_1}$ (in (\ref{eq:g_freq_conv})) is non-zero, then $o(\sqrt{\abs{E_n}})$ can be replaced by $o(\lambda_{t,n})$. Hence for large $n$, we have $\lambda_{t,n} \approx k_t\sqrt{\abs{E_n}}$ and it is reasonable to fit points of the form $(\sqrt{\abs{E_n}}, \lambda_{t,n})$ by a line passing through the origin $(0,0)$.

On the other hand, by the same argument, if $(G_n)_{1\leq n\leq 90}$ converges to a \emph{non-zero} classical graphon in the cut distance, then according to \cite[Lemma 4]{Lua21}, points of the form $(|V_n|, \lambda_{t,n})$ can also be approximately fitted by a line through $(0,0)$ for large $n$. Finally, if $(G_n)_{1\leq n\leq 90}$ has bounded degree as assumed in \cite{RodGamBar:J22}, called ``Graphing'', then it is known that $\set{\lambda_{t,n}\given 1\leq n\leq 90}$ is bounded \cite{Lov07}. In this case, for each tail fit, we use a horizontal line.

To verify which model is suitable for the data, we consider linear fittings as described above to our model, the classical graphon model \cite{Lua21}, and the ``Graphing'' model \cite{RodGamBar:J22}, respectively. We use the tail parts of the sequences, i.e., $G_n$ with $n\geq 64$ ($|V|_{64} \approx 6600$), $n\geq 76$ ($|V|_{76} \approx 8000$), and $n\geq 82$ ($|V|_{82} \approx 8600$) respectively, for the fittings. We then calculate the mean squared error (MSE) of the fitted lines. 

From \cref{tab:MSE}, we see that the linear model for $(\sqrt{\abs{E_n}}, \lambda_{t,n})$ fits better than that for $(\abs{V_n}, \lambda_{t,n})$. The underperformance of the classical approach is due possibly to the sparsity of the graph sequence (see \cref{fig:edl,fig:tnd}), and the true limiting object is the zero graphon (see the next paragraph). In addition, contrary to \cite{RodGamBar:J22}, the magnitudes of the eigenvalues increase instead of being clearly bounded by some constant. It also has a worse fit as graph sequence with bounded degree is ``too sparse''. 

\begin{figure}[!htb]
\centering
\begin{subfigure}[b]{0.7\columnwidth}
\centering
\includegraphics[width=\linewidth, trim={0cm 0cm 0 0cm},clip]{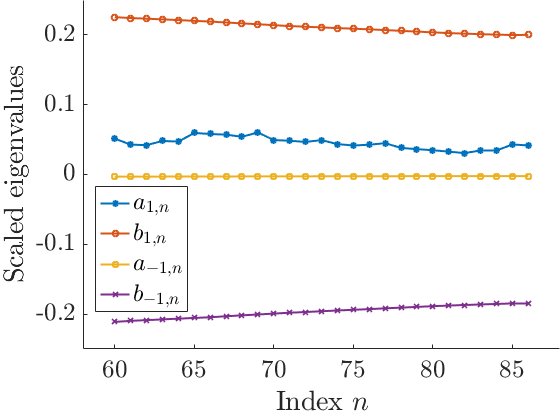}
\end{subfigure}
\centering
\begin{subfigure}[b]{0.7\columnwidth}
\centering
\includegraphics[width=\linewidth, trim={0cm 0cm 0 0cm},clip]{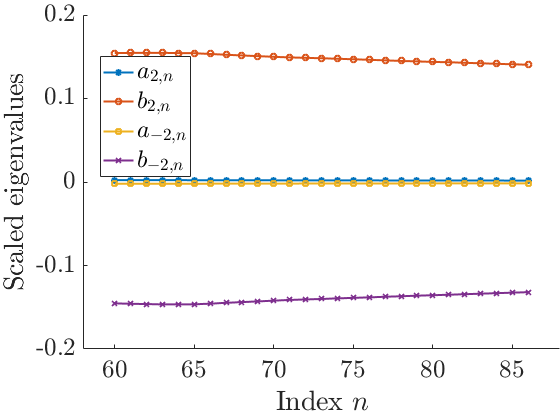}
\end{subfigure}
\centering
\begin{subfigure}[b]{0.7\columnwidth}
\centering
\includegraphics[width=\linewidth, trim={0cm 0cm 0 0cm},clip]{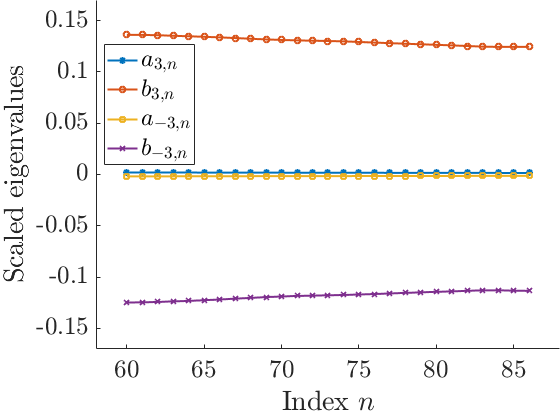}
\end{subfigure}
\caption{The figures are the plots for $a_{t,n}, b_{t,n}$ as averages of eigenvalues scaled by $|V_n|, \sqrt{|E_n|}$ respectively, with $-3\leq t\leq 3, 60\leq n\leq 86$.} \label{fig:tpf}
\end{figure}

To provide evidence that $G_n$ might converge to the zero graphon in the classical graphon sense, we compute and plot the average $a_{t,n}$ of $\{\lambda_{t,n}/|V_n|,\ldots, \lambda_{t,n+4}/|V_{n+4}|\}$ for $60 \leq n \leq 86, -3\leq t\leq 3$. For comparison, we also obtain the average $b_{t,n}$ of $\{\lambda_{t,n}/\sqrt{|E_n|},\ldots, \lambda_{t,n+4}/\sqrt{|E_{n+4}|}\}$ from the generalized graphon theory. As shown in \cref{fig:tpf}, both $(a_{t,n})_{60 \leq n \leq 86}$ and $(b_{t,n})_{60 \leq n \leq 86}$ generally display (horizontal) convergence trend as $n$ increases. However, for most $t$ and $n$, each $a_{t, n}$ is very close to $0$, which supports our earlier speculation on the vanishing of the limiting graphon. 

In summary, generalized graphons are worthy limiting objects to consider if one wants to avoid a trivial limit.

\subsection{Cora: convergence of filter coefficients}

\begin{figure}[!htb]
\centering
\includegraphics[scale=0.3]{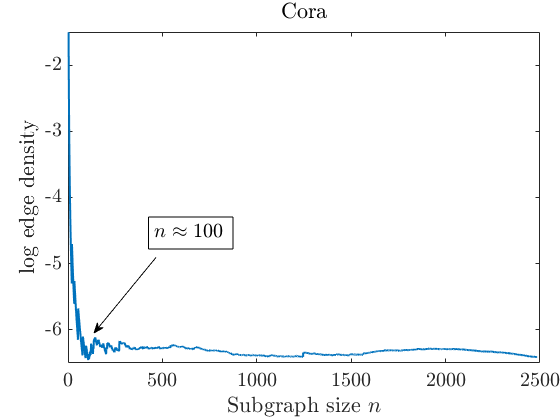}
\caption{Edge densities ($\log$) of (growing) subgraphs of the Cora graph.} 
\label{fig:edo}
\end{figure}

\begin{figure}[!htb]
    \centering
    \includegraphics[scale=0.3]{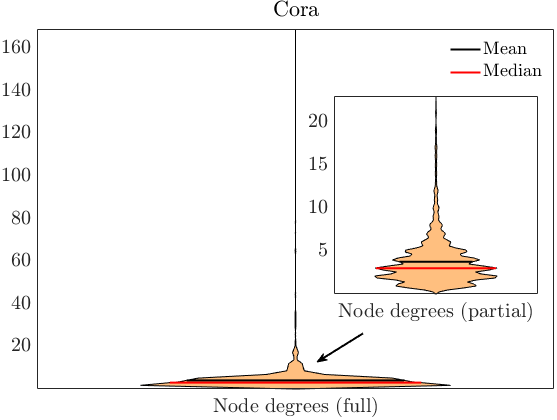}
    \caption{For the node degree distribution of the Cora graph, we show both the full distribution and the partial distribution where $1\%$ nodes with the largest degrees are removed. We again see a $\log$-tailed pattern.}
    \label{fig:tndd}
\end{figure}

In this study, we consider the Cora graph $G=(V,E)$ of size $2485$.\footnote{https://graphsandnetworks.com/the-cora-dataset/} The graph size is smaller than the Ogbn-arXiv graph, and we adopt a simpler sampling strategy. We randomly order the nodes to model the random sample procedure and obtain a sequence of growing graphs $(G_m)_{1\leq m\leq 2485}$ with $G_{2485}=G$ the full graph. The edge densities (in $\log$ scale) of this sequence of graphs are shown in \cref{fig:edo}. We see that $(G_m)_{1\leq m\leq 2485}$ is indeed a sparse graph sequence with edge densities decaying quickly. Its degree distribution in \cref{fig:tndd} suggests that the graph has a dense core (cf.\ \cref{eg:cas}), while the overall pattern is log-tailed. 

\begin{figure}[!htb]
\centering
\includegraphics[scale=0.3]{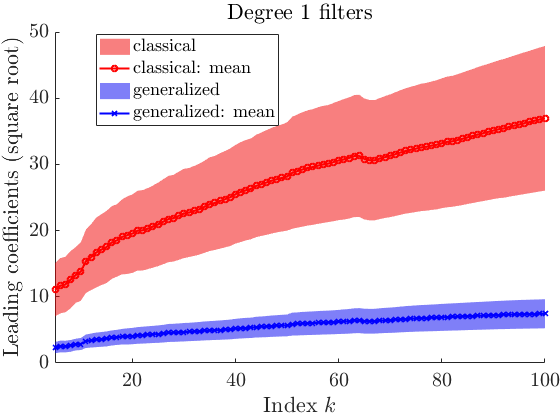}
\includegraphics[scale=0.3]{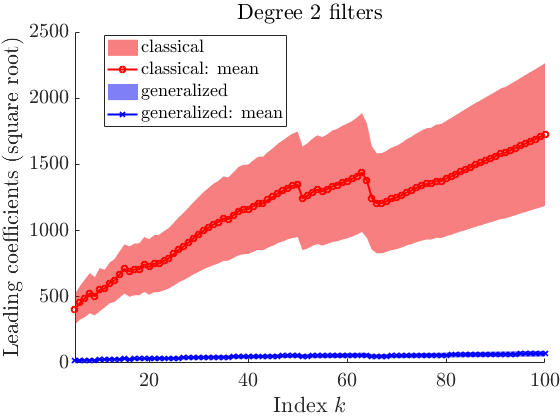}
\caption{The (square root) sequences of coefficients $(c^1_{d,k})_{60\leq k\leq 100}$ and $ (c^2_{d,k})_{60\leq k\leq 100}$. The shaded standard deviation regions are from $40$ runs.}
\label{fig:tso}
\end{figure}

\begin{figure}[!htb]
\centering
\includegraphics[scale=0.33]{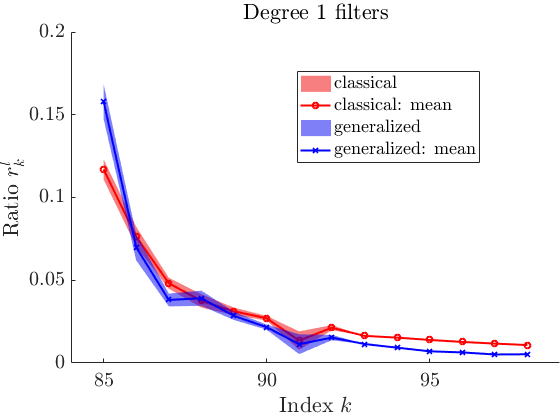}
\includegraphics[scale=0.33]{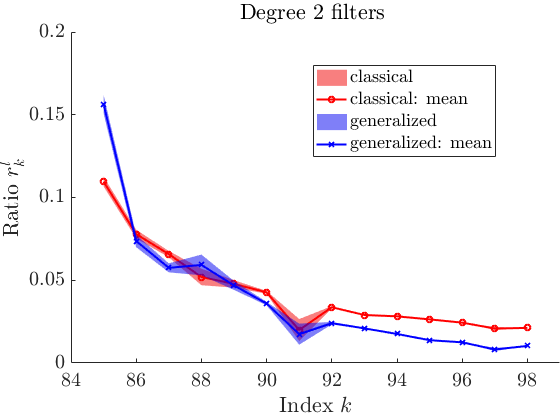}
\caption{The plots of $r_k^1$ and $r_k^2$ for $k\geq 85$.}
\label{fig:tpo}
\end{figure}

To study filter convergence, we choose a subsequence $(G_{m_k})_{1\leq m_k\leq 100}$ of size $100$ where $m_k = 24k$ and $G_{m_{100}}=G$. We synthesize graph signals on $G$ by the following procedure that mimics information diffusion on the graph: 
\begin{itemize}
\item On $10\%$ of nodes with the largest degrees, we generate independent uniform random numbers in $[0,1]$. On each remaining node, the signal is $0$. The resulting graph signal is denoted by $f$.   
\item Let $A_G$ be the adjacency matrix of $G$. We apply a polynomial filter 
\begin{align*}
P_d(A_G) = \sum_{0\leq i\leq d}c_iA_G^i
\end{align*}
in $A_G$ of degree $d$, with uniform random coefficients $c_0,\ldots,c_d$ in $[0,1]$, to $f$. This is a $d$-step diffusion on $G$. The resulting signal is denoted by $g$. 
\end{itemize}

On each subgraph $G_{m_k}$, we obtain the corresponding source signal $f_k$ and target signal $g_k$. Using simple regression, we want to estimate a polynomial filter that transforms $f_k$ to $g_k$. For the purpose of this paper, two approaches are compared here: ``classical'' and ``generalized''. The ``classical'' method is based on \cite{Lua21, JianJiTay:J23}, a polynomial in $A_{G_{m_k}}/m_k$ is estimated. On the other hand, let $e_k$ be the edge size of $G_{m_k}$. The approach ``generalized'' is based on the current work with a polynomial in $A_{G_{m_k}}/\sqrt{2e_k}$ being estimated. As the leading term has the largest impact, we investigate the convergence behavior of the sequences of estimated leading coefficients $(c^1_{d,k})_{k\leq 100}$ for ``classical'' and $(c^2_{d,k})_{k\leq 100}$ for ``generalized''. For $k=100$, we obtain the ground truth, also considered as the ``limit''.

The objective of this study is to determine the convergence behavior of $(c^1_{d,k})_{k\leq 100}$ and $(c^2_{d,k})_{k\leq 100}$. We show the square roots of the sequences so that they can be compared in the same plot. In each plot of \cref{fig:tso}, we see that ``generalized'' displays a more steady convergence to the limit and suffers less from exploding coefficients.  

To quantify for the convergence trend of the sequences, let $\overline{c_d^1}$ (resp.\ $\overline{c_d^2}$) be the average of $(c^1_{d,k})_{85\leq k\leq 100}$ (resp.\ $(c^2_{d,k})_{85\leq k\leq 100}$). 
We compute the ratio
\begin{align*}
r_k^l = \frac{|c_{d,k+1}^l-c_{d,k}^l|}{|c_{d,100}^l-\overline{c_d^l}|},\ l=1,2,
\end{align*}
as a measure of the relative error change for each $k$. The smaller its value at the tail indicating a relative smaller change of the estimated coefficients at the tail, implying a faster convergence. Plots in \cref{fig:tpo} agree with our observation in \cref{fig:tso} that ``generalized'' converges faster. 

\section{Conclusion} \label{sec:con}
We have presented a signal processing framework based on the theory of generalized graphons, which include limit objects of sparse sequences of graphs with respect to the stretched cut distance. This framework is suitable for various types of large real-world networks. We have obtained operator convergence results for sequences of graphons that converge in the stretched cut distance. Additionally, we have proposed a sampling scheme for generalized graphons and demonstrated convergence results for sampled graphs, their spectra, and their signals.

An interesting future work is to extend this signal processing framework to incorporate the graphex model \cite{Bor18, Bor19}. The graphex model includes not only the "graphon component" discussed in this paper but also the "star component" and the "dust component," which represent star subgraphs and isolated edges, respectively. These components collectively describe the "non-core parts" of graphs. By incorporating the graphex model, we aim to further enhance our signal processing framework for a broader range of graph structures.

\appendices

\section{Proofs of theoretical results} \label{app:pot}

\begin{IEEEproof}[Proof of \cref{lem:nlf}]
We apply H\"{o}lder's inequality:
\begin{align*}
&\norm{T_{W^{\mathfrak{s}}}f}_2^2 = \int_{\mathbb{R}_+}\abs*{\int_{\mathbb{R}_+}W^{\mathfrak{s}}(x,y)f(y)\ud y}^2\ud x \\
&\leq \int_{\mathbb{R}_+}\parens*{\int_{\mathbb{R}_+}|W^{\mathfrak{s}}(x,y)|\cdot |f(y)|\ud y}^2\ud x \\
&\leq \int_{\mathbb{R}_+} \int_{\mathbb{R}_+}|W^{\mathfrak{s}}(x,y)|^2 \ud y \cdot \int_{\mathbb{R}_+} |f(y)|^2\ud y \ud x \\
&= \norm{f}_2^2\int_{\mathbb{R}_+^2} |W^{\mathfrak{s}}(x,y)|^2\ud y \ud x \\ 
&= \norm{W^{\mathfrak{s}}}_2^2\norm{f}_2^2
= \frac{\norm{W}_2^2}{\norm{W}_1^2}\norm{f}_2^2.
\end{align*}
H\"{o}lder's inequality is used to derive the second inequality for the estimation of the inner integral.    
\end{IEEEproof}

\begin{IEEEproof}[Proof of \cref{thm:iwi}]
As the graphons are assumed to be uniformly bounded, say by $b$, then by replacing the sequence by $(W_n/b)_{n> 0}$ if necessary, we may assume that $b\leq 1$. The strategy of the proof is to consider separately the dominant part (in $2$-norm) and tail part of each $W_n$, as the former is compactly supported and the latter is small in size ($2$-norm). 

By \cite[Theorem 15]{Bor18}, the sequence $(W_n^{\mathfrak{s}})_{n>0}$ has uniformly regular tails. This means that for any $\epsilon_1 > 0$, there is a subset $U \subset \mathbb{R}_+$ with finite Lebesgue measure $\mu(U) = M$ such that (using $b\leq 1$): 
\begin{align}
\norm{W_n^{\mathfrak{s}} - W_n^{\mathfrak{s}}1_{U\times U}}_2 \leq \norm{W_n^{\mathfrak{s}} - W_n^{\mathfrak{s}}1_{U\times U}}_1  \leq \epsilon_1 \label{ineq:ws}
\end{align} 
for $n> 0$, where $1_{U\times U}$ is the indicator function over $U\times U$. By enlarging $U$ if necessary, we further assume that $M\geq 1$ and the same inequalities hold for $W^{\mathfrak{s}}$ (in place of $W_n^{\mathfrak{s}}$). 

As $\mu(U)=M<\infty$, we may endow $U$ with the uniform probability measure. The purpose of doing so is that we want to apply \cite[Lemma E.6]{Jan:J13}. We next need to find the relations between different norms for different measures. Let $\norm{\cdot}_{2,U}$ be the $2$-norm on $U$ with the uniform probability measure. Then for any $f \in L^2(\mathbb{R}_+)$ supported on $U$, we have $\norm{f}_2 = M\norm{f}_{2,U}$.
We use $\norm{\cdot}_{2,2,U}$ and $\norm{\cdot}_{\square,U}$ to denote the operator norm and cut norm on $L^2(U)$ with the uniform probability measure. Then $M^2\norm{\cdot}_{\square,U} = \norm{\cdot}_{\square}$. On the other hand, if $T: L^2(U) \to L^2(U)$ is a bounded linear operator, then 
\begin{align*}
\norm{Tf}_{2} & = M\norm{Tf}_{2,U} \leq M\norm{T}_{2,2,U} \norm{f}_{2,U} \\
& = \norm{T}_{2,2,U} \cdot \norm{f}_{2}.
\end{align*}
Hence, if $\norm{\cdot}_{2,2}$ denotes the operator norm of $T$ \gls{wrt} the Lebesgue measure $\mu$, then we have 
\begin{align}\label{ineq:2-2}
\norm{\cdot}_{2,2} \leq \norm{\cdot}_{2,2,U}.
\end{align}
Let $W_n' = W_n^{\mathfrak{s}}1_{U\times U}$ and $W' = W^{\mathfrak{s}}1_{U\times U}$. 
We have 
\begin{align*} 
& \norm{W_n'- W'}_{\square} \leq \norm{W_n^{\mathfrak{s}}-W^{\mathfrak{s}}}_{\square}\\
& \qquad + \norm{(W_n^{\mathfrak{s}} - W_n')-(W^{\mathfrak{s}} - W')}_{\square} \\
&= d_{\square,\mathfrak{s}}(W_n, W)\\ 
&\qquad + \norm{(W_n^{\mathfrak{s}} - W_n^{\mathfrak{s}}1_{U\times U})-(W^{\mathfrak{s}} - W^{\mathfrak{s}}1_{U\times U})}_{\square} \\
&\leq d_{\square,\mathfrak{s}}(W_n, W) + \norm{W_n^{\mathfrak{s}} - W_n^{\mathfrak{s}}1_{U\times U}}_{\square} \\ 
&\qquad +\norm{W^{\mathfrak{s}} - W^{\mathfrak{s}}1_{U\times U}}_{\square} \\
&\leq d_{\square,\mathfrak{s}}(W_n, W) + \norm{W_n^{\mathfrak{s}} - W_n^{\mathfrak{s}}1_{U\times U}}_1\\ 
&\qquad +\norm{W^{\mathfrak{s}} - W^{\mathfrak{s}}1_{U\times U}}_
1\\
&\leq  d_{\square,\mathfrak{s}}(W_n, W) + 2\epsilon_1. \numberthis \label{eq:nln}
\end{align*}
By \cref{ineq:2-2} comparing $\norm{\cdot}_{2,2}$ and $\norm{\cdot}_{2,2,U}$, we have
\begin{align*}
& \norm{W_n'-W'}_{2,2} \leq \norm{W_n'-W'}_{2,2,U} \\
&\leq  \sqrt{2}\norm{W_n'-W'}_{\square,U}^{1/2} =\frac{\sqrt{2}}{M}\norm{W_n'-W'}_{\square}^{1/2} \\
&\leq \sqrt{2}\norm{W_n'-W'}_{\square}^{1/2}, \numberthis\label{eq:lnl}
\end{align*}
where the second inequality follows from \cite[Lemma E.6]{Jan:J13}.

Since we assume that $W_n \to W$ in $d_{\square,\mathfrak{s}}$ as $n\to \infty$, for any $\epsilon_2 >0$, we can choose $N$ such that if $n>N$, $d_{\square,\mathfrak{s}}(W_n, W)\leq \epsilon_2$. Hence, substituting \cref{eq:nln} in \cref{eq:lnl},
we obtain $\norm{W_n'-W'}_{2,2} \leq \sqrt{2}(\epsilon_2+2\epsilon_1)^{1/2}$. Now for $f \in L^2(\mathbb{R})$, we have
\begin{align*}
\norm{T_{W_n'}f-T_{W'}f}_2 \leq \sqrt{2}(\epsilon_2+2\epsilon_1)^{1/2}\norm{f}_2.
\end{align*}
On the other hand, by the same argument as in the proof of \cref{lem:nlf} using H\"{o}lder's inequality, we have 
\begin{align*}
\norm{T_{W_n^{\mathfrak{s}}}f-T_{W_n'}f}_2^2 
\leq \norm{f}_2^2\norm{W_n^{\mathfrak{s}}-W_n^{\mathfrak{s}}1_{U\times U}}_2^2
\leq \norm{f}_2^2\epsilon_1^2,
\end{align*}
where the last inequality follows from \cref{ineq:ws}. 
Similarly, $\norm{T_{W^{\mathfrak{s}}}f-T_{W'}f}_2 \leq \epsilon_1\norm{f}_2$. Then, by the triangle inequality, we have 
\begin{align*}
&\norm{T_{W_n^{\mathfrak{s}}}f-T_{W^{\mathfrak{s}}}f}_2 \\
&\leq \norm{T_{W_n'}f-T_{W'}f}_2 + \norm{T_{W_n^{\mathfrak{s}}}f-T_{W_n'}f}_2\\
&\qquad + \norm{T_{W^{\mathfrak{s}}}f-T_{W'}f}_2\\
&\leq (\sqrt{2}(\epsilon_2+2\epsilon_1)^{1/2}+2\epsilon_1)\norm{f}_2.
\end{align*}
The result follows from first choosing $U$ so that $\epsilon_1$ is small enough and then choosing $N$ so that $\epsilon_2$ is sufficiently small. 

If $W_n \to W$ in $\delta_{\square,\mathfrak{s}}$, then by \cite[Proposition 48]{Bor18}, there is a sequence of measure preserving $\phi_n$ such that ${W_n^{\mathfrak{s}}}^{\phi_n} \to W^{\mathfrak{s}}$ in $d_{\square,\mathfrak{s}}$. Hence, the claim follows from the first part.
\end{IEEEproof}

\begin{IEEEproof}[Proof of \cref{coro:lpb}]
By the triangle inequality, it suffices to prove the statement when $P$ is a monomial of the form $P(x) = x^k$. The case $k=0$ is trivial and $k=1$ follows from \cref{thm:iwi}. We proceed by induction for $k>1$. As $n\to \infty$, we have
\begin{align*}
& \norm{T_{W_n^{\mathfrak{s}}}^k - T_{W^{\mathfrak{s}}}^k}_{2,2} \\ 
& \leq \norm{T_{W_n^{\mathfrak{s}}}\circ (T_{W_n^{\mathfrak{s}}}^{k-1} - T_{W^{\mathfrak{s}}}^{k-1})}_{2,2}  + \norm{(T_{W_n^{\mathfrak{s}}} - T_{W_n^{\mathfrak{s}}})\circ T_{W^{\mathfrak{s}}}^{k-1}}_{2,2}.
\end{align*}
The right-hand-side converges to $0$, since 
\begin{align*}
\norm{T_{W_n^{\mathfrak{s}}}^{k-1} - T_{W^{\mathfrak{s}}}^{k-1}}_{2,2} \to 0 \text{ and } \norm{T_{W_n^{\mathfrak{s}}} - T_{W^{\mathfrak{s}}}}_{2,2} \to 0
\end{align*}
by the induction hypothesis.
\end{IEEEproof}

\begin{IEEEproof}[Proof of \cref{coro:swi}]
We have
\begin{align*}
& \norm{T_{W_n^{\mathfrak{s}}}f_n - T_{W^{\mathfrak{s}}}f}_2 \\
&\leq  \norm{T_{W_n^{\mathfrak{s}}}f_n - T_{W^{\mathfrak{s}}}f_n}_2 + \norm{T_{W^{\mathfrak{s}}}f_n - T_{W^{\mathfrak{s}}}f}_2 \\
&\leq \norm{T_{W_n^{\mathfrak{s}}} - T_{W^{\mathfrak{s}}}}_{2,2}\norm{f_n}_2 + \norm{W}_{2,2,\mathfrak{s}}\norm{f_n-f}_2.
\end{align*}
Since $f_n \to f$ in $L^2(\mathbb{R}_+)$ as $n \to \infty$, we have $\norm{f_n}_2$ is bounded for all $n$. Both $\norm{T_{W_n^{\mathfrak{s}}} - T_{W^{\mathfrak{s}}}}_{2,2}$ (\cref{thm:iwi}) and $\norm{f_n-f}_2$ converge to $0$ as $n\to \infty$, so does $\norm{T_{W_n^{\mathfrak{s}}}f_n - T_{W^{\mathfrak{s}}}f}_2$. 
\end{IEEEproof}

To prove \cref{thm:swi}, we need a few lemmas.

\begin{Lemma} \label{lem:ina}
If non-negative and integrable graphons $W_n$ converge to $W$ in the cut distance $\delta_{\square}$, then $\norm{W_n}_1 \to \norm{W}_1$ as $n\to \infty$.
\end{Lemma}
\begin{IEEEproof}
For any measure preserving functions $\phi_1,\phi_2$, we have the following:
\begin{align*}
\abs*{\norm{W_n}_1 - \norm{W}_1} 
& =  \abs*{\int_{\mathbb{R}_+^2} \parens*{W_n^{\phi_1}(x,y) - W^{\phi_2}(x,y)} \ud x\ud y}\\ 
& \leq d_{\square}(W_n^{\phi_1},W^{\phi_2}).
\end{align*}
Therefore, $\abs{\norm{W_n}_1 - \norm{W}_1} \leq \delta_{\square}(W_n,W)$ and the result follows.
\end{IEEEproof}

\begin{Lemma} \label{lem:ifo}
Suppose $f$ on $\mathbb{R}_+$ or $\mathbb{R}_+^2$ is bounded, compactly supported and Riemann integrable. If a positive sequence $r_n \to r > 0, n \to \infty$, then $f^{r_n} \to f^r$ in $1$-norm, where $f^{r_n}(x) = f(r_nx)$ is the function stretched by $r_n$. 
\end{Lemma}
\begin{IEEEproof}
We prove the case where $f$ is a function on $\mathbb{R}_+$ and the proof for $f$ being a function on $\mathbb{R}_+^2$ is identical. Let $[0,t]$ cover the support of $f$ and $b>0$ be a bound of $f$.

For $\epsilon>0$, as $f$ is Riemann integrable, we may subdivide $[0,t]$ into $n_0$ equal non-overlapping subintervals $\{I_l, 1\leq l\leq n_0\}$ and form a Riemann sum $R$ such that $\norm{f-R}_1 < \epsilon$. Here, $R$ is a step function constant on each $I_l$. We obtain
\begin{align*}
& \norm{f^{r_n}-f^r}_1 \\ & \leq \norm{f^{r_n}-R^{r_n}}_1 + \norm{f^r-R^r}_1+\norm{R^{r_n}-R^r}_1.
\end{align*}
Apparently, $\norm{f^{r_n}-R^{r_n}}_1 < \epsilon/r_n$ and $\norm{f^r-R^r}_1 < \epsilon/r$. On the other hand, 
\begin{align*}
\norm{R^{r_n}-R^r}_1 \leq n_0\cdot (2b)\cdot t|\frac{1}{r_n}-\frac{1}{r}|  = 2btn_0|\frac{1}{r_n}-\frac{1}{r}|.
\end{align*}
This is because each of the $n_0$ intervals of the supports of $R^{r_n}$ and $R^r$ differ at most by $n_0\cdot(t/n_0)|1/r_n-1/r| = t|1/r_n-1/r|$ in size (illustrated in \cref{fig:rrn}); and for each $x$ on the difference, $|R^{r_n}(x)-R^r(x)| \leq 2b$. 
 In summary, we have 
\begin{align*}
\norm{f^{r_n}-f^r}_1 \leq (\frac{1}{r_n}+\frac{1}{r})\epsilon + 2btn_0|\frac{1}{r_n}-\frac{1}{r}|.
\end{align*}
As $r>0$, $1/r_n \to 1/r$ as $n\to \infty$. Therefore, if we first choose $n_0$ be sufficiently large, and then let $n$ be large enough such that $n_0|1/r_n-1/r|$ is sufficiently small, then $\norm{f^{r_n}-f^r}_1$ can be arbitrarily small. Hence, $f^{r_n} \to f^r$ in $1$-norm.
\begin{figure}
\centering
\includegraphics[scale=0.55]{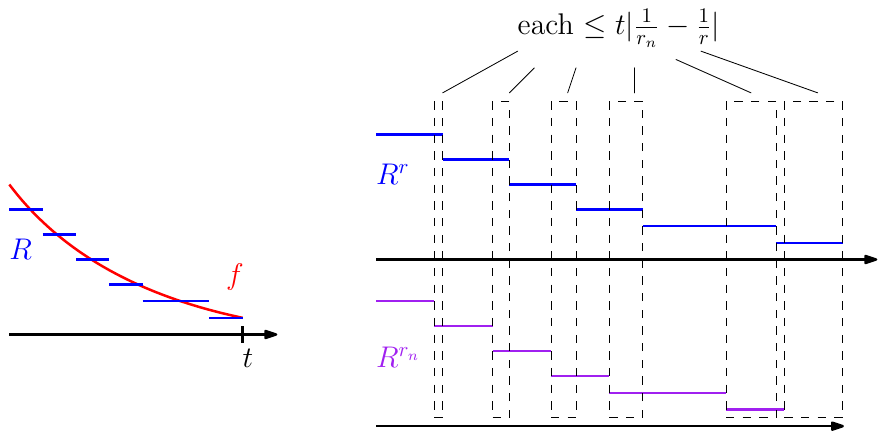}
\caption{Illustration of the difference between $R^r$ and $R^{r_n}$.} 
\label{fig:rrn}
\end{figure}
\end{IEEEproof}

Recall that $(t_m)_{m>0}$ is a strictly increasing sequence of positive real numbers such that $t_m \to \infty$ as $m \to \infty$. 
 
\begin{Lemma} \label{lem:fta}
Following the assumptions of \cref{thm:swi}, we have $W_m\stret \to W\stret$ in $1$-norm as $m\to \infty$.   
\end{Lemma}

\begin{IEEEproof}
Write $W_m^{\mathfrak{s}}$ (stretched \gls{wrt} $r_m=\norm{W_m}_1^{1/2}$) as $W_m^{r_m}$, and $W^{\mathfrak{s}}$ (stretched \gls{wrt} $r=\norm{W}_1^{1/2}$) as $W^r$. 

From the dominated convergence theorem, $W_m$ converges to $W$ in $1$-norm as $m\to \infty$. 
Therefore, for any $\epsilon_1>0$, we can choose $m_0$ such that $\norm{W-W_{m_0}}_1\leq \epsilon_1$. The graphon $W_{m_0}$ is compactly supported. For any $m \geq m_0$, we have
\begin{align*}
& \norm{W^{\mathfrak{s}}-W_m^{\mathfrak{s}}}_1 = \norm{W^r-W_m^{r_m}}_1\\
& = \norm{W_{m_0}^r+(W-W_{m_0})^r-W_{m_0}^{r_m}-(W_m-W_{m_0})^{r_m}}_1\\
&\leq \norm{W_{m_0}^r-W_{m_0}^{r_m}}_1 + \norm{(W-W_{m_0})^r}_1 \\& \qquad + \norm{(W_m-W_{m_0})^{r_m}}_1 \\
&\leq \norm{W_{m_0}^r-W_{m_0}^{r_m}}_1 + \epsilon_1\parens*{\frac{1}{r^2}+\frac{2}{r_m^2}}.
\end{align*}
As $m\to \infty$, we have $r_m\to r$. By \cref{lem:ifo}, the first term converges to $0$ as $m \to \infty$. Therefore, if we first let $\epsilon_1$ to be sufficiently small and then let $m$ to be sufficiently large, $\norm{W^{\mathfrak{s}}-W_m^{\mathfrak{s}}}_1$ can be made arbitrarily small. This proves that $W_m^{\mathfrak{s}} \to W^{\mathfrak{s}}$ in $1$-norm as claimed.
\end{IEEEproof}

\begin{IEEEproof}[Proof of \cref{thm:swi}]
We want to show that $W_{m,n} \to W$ in the stretched cut distance $\delta_{\square,\mathfrak{s}}$ as $m,n\to\infty$ by the following steps:
\begin{enumerate}[(a)]
\item Recall $W_m=W1_{[0,t_m]^2}$ defined in (\ref{eq:www}). We first show that with probability one, $W_{m,n} \to W_m$ in the stretched cut distance $\delta_{\square,\mathfrak{s}}$, or equivalently, $W_{m,n}\stret \to W_m\stret$ in the cut distance $\delta_{\square}$, as $n\to\infty$ for each $m>0$. 
\item Then, we show that $W_m\stret \to W\stret$ in $1$-norm as $m\to \infty$.
\end{enumerate}
Since the $1$-norm dominates the cut norm, the theorem follows. 

As (b) is already obtained in \cref{lem:fta}, it remains to prove (a). For this, we transform the problem so that known results for classical graphon convergence can be applied. This is essentially because the stretched and classical cut norms are equivalent on finite intervals.

It is obvious that $W_m$ is supported on $[0,t_m]^2$. Recall in (\ref{eq:www}), $W_m'$ is the graphon on $[0,1]^2$ defined by 
\begin{align*}
W_m'(x,y) = W_m(t_mx,t_my),
\end{align*}
i.e, $W_m'$ is $W_m$ stretched (or shrunk) by $t_m$ to be supported on $[0,1]^2$ (see \cref{fig:wax}). Moreover, $\norm{W_m}_1 = t_m^2\norm{W_m'}_1$. Thus, for $W_m^{\mathfrak{s}}$ (\gls{wrt} $\norm{W_m}_1^{1/2}$) and $W_m'^{\mathfrak{s}}$ (\gls{wrt} $\norm{W_m'}_1^{1/2}$), we have
\begin{align} \label{eq:wmw}
W_m^{\mathfrak{s}} = W_m'^{\mathfrak{s}}.
\end{align}

\begin{figure}
\centering
\includegraphics[scale=0.55]{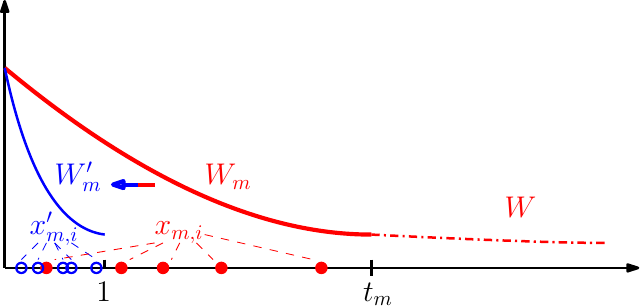}
\caption{An illustration of $W_m, W_m'$ and $\{x_{m,i},1\leq i\leq n\}$. For $W_m$ and $W_m'$, only a cross-section whose domain is $\mathbb{R}_+$ is shown.} 
\label{fig:wax}
\end{figure}

The graph $G_{m,n}$ can be equivalently obtained by sampling uniformly random independent points $x_{m,1}',\ldots, x_{m,n}'$ from $[0,1]$ and forming an edge $(v_{m,i}, v_{m,j})$ with probability 
\begin{align*}
W_m'(x_{m,i}',x_{m,j}') = W_m(t_mx_{m,i}',t_mx_{m,j}').
\end{align*}
By \cite[Lemma 10.16]{Lov:12}, as $n\to \infty$, $W_{m,n}$ converges to $W_m'$ in the cut distance $\delta_{\square}$. Therefore, by \cref{lem:ina}, 
\begin{align} \label{eq:wm1}
\norm{W_{m,n}}_1 \to \norm{W_m'}_1 \text{ as } n \to \infty. 
\end{align}
Let $W_m^{n,\mathfrak{s}}$ be the stretch of $W_m'$ \gls{wrt} $\norm{W_{m,n}}_1^{1/2}$. We have
\begin{align*}
\delta_{\square}(W_{m,n}^{\mathfrak{s}}, W\stret_m) 
&= \delta_{\square}(W_{m,n}^{\mathfrak{s}}, W_m'^{\mathfrak{s}}) \\
& \leq \delta_{\square}(W_{m,n}^{\mathfrak{s}},W_m^{n,\mathfrak{s}}) + \norm{W_m^{n,\mathfrak{s}} - W_m'^{\mathfrak{s}}}_1. 
\end{align*}
As $n\to\infty$, the term $\delta_{\square}(W_{m,n}^{\mathfrak{s}},W_m^{n,\mathfrak{s}}) \to 0$ since $W_{m,n} \to W_m$ in the cut distance 
and $\norm{W_{m,n}}_1$ is bounded. Moreover, $\norm{W_{m,n}}_1$ is lower bounded away from $0$ for $m,n$ large enough. 
On the other hand, $\norm{W_m^{n,\mathfrak{s}} - W_m'^{\mathfrak{s}}}_1$ converges to $0$ by \cref{lem:ifo}. Hence, $\delta_{\square}(W_{m,n}^{\mathfrak{s}}, W_m'^{\mathfrak{s}})\to 0$ as $n\to \infty$. This proves (a) and hence the theorem.
\end{IEEEproof}

\begin{IEEEproof}[Proof of \cref{coro:ted}]
Following the proof of \cref{thm:swi} (cf.\ notations (\ref{eq:www})), with probability one, the limit edge density of $(G_{m,n})_{n\geq 1}$ (with fixed $m$) is
\begin{align*}
    \norm{W_m'}_1 = \frac{\norm{W_m}_1}{t_m^2} \leq \frac{\norm{W}_1}{t_m^2}.
\end{align*}
Since $\norm{W}_1 <\infty$ and $t_m\to\infty$ as $m\to\infty$, we have $\lim\limits_{m\to \infty}\lim\limits_{n\to \infty}\calE(G_{m,n})=0$.
\end{IEEEproof}

\begin{IEEEproof}[Proof of \cref{thm:sfi}]
For each $m>0$, similar to $W_m, W_m'$ introduced in \cref{eq:www}, define $f_m = f1_{[0,t_m]}$ and $f_m'$ such that $f_m'(x) = f_m(t_mx)$. These functions are supported on $[0,t_m]$ and $[0,1]$ respectively. By the same argument as the proof of \cref{thm:swi} (using \cref{lem:ifo}), we have $f_m^{\mathfrak{s}} \to f^{\mathfrak{s}}$ in $1$-norm. Therefore, we want to show that with probability one, $f_{m,n}^{\mathfrak{s}} \to f_m^{\mathfrak{s}}$ in $1$-norm. Notice that $f_m'$ stretched w.r.t.\ $r_m'=\norm{W_m'}_1^{1/2}$ is the same as $f_m^{\mathfrak{s}}$. We have 
\begin{align*}
    & \norm*{f_{m,n}^{\mathfrak{s}}-f_m^{\mathfrak{s}}}_1 = \norm*{f_{m,n}^{r_{m,n}}-{f_m'}^{r_m'}}_1 \\
    & \leq \norm*{f_{m,n}^{r_{m,n}}-{f_m'}^{r_{m,n}}}_1 + \norm*{{f_m'}^{r_{m,n}}-{f_m'}^{r_m'}}_1.
\end{align*}
From \cref{eq:wm1}, $r_{m,n}=\norm{W_{m,n}}_1^{1/2} \to r_m'$ as $n\to\infty$. 
Applying \cref{lem:ifo}, we obtain $\norm{{f_m'}^{r_{m,n}}-{f_m'}^{r_m'}}_1 \to 0$. As $(r_{m,n})_{n>0}$ is bounded, it suffices to show that with probability one, $f_{m,n}$ converges in $1$-norm to the \gls{aects} function $f_m'$. 

We construct a graph parameter $\rho_m$ \cite[Section 10.2]{Lov:12} as follows. For the (random graph) $G_{m,n}$, let 
\begin{align*}
\rho_m(G_{m,n}) = \norm{f_{m,n}-f_m'}_1.
\end{align*}
As $f_{m,n}$, and hence $n^2\rho_m$, depends only on the vertex set of the graph, they satisfy the reasonable smoothness condition of \cite[p.159]{Lov:12}. Applying the concentration inequality \cite[Theorem 10.2]{Lov:12}  to $n^2\rho_m$ (setting $t =k=n$), we have:  
\begin{align*}
    \mathbb{P}\Big(\rho_m(G_{m,n})\geq \E[\rho_m(G_{m,n})] + \sqrt{2}n^{-1}\Big)\leq \exp(-n). 
\end{align*}
Notice $\sum_{n>0} \exp(-n) < \infty$. By the Borel–Cantelli Lemma, with probability one, $\rho_m(G_{m,n})< \E[\rho_m(G_{m,n})] + \sqrt{2}n^{-1}$ for all except finitely many $n$. 
Hence, to prove the theorem, it suffices to show that as $n\to \infty$, $\E[\rho_m(G_{m,n})]=0$.

We have seen in the proof of \cref{thm:swi} that $G_{m,n}$ can be obtained by sampling $S_{m,n}=\{x_{m,1}',\ldots, x_{m,n}'\}$ uniformly and independently from $[0,1]$ and forming edges using $W_m'$. Similarly, $f_{m,n}$ can be obtained as a step function by evaluating $f_m'$ at $S_{m,n}$ at each step, where a step has a size of $1/n$. 

Consider an integer $k>0$. Subdivide $[0,1]$ into $k$ equal subintervals $I_l$, $1\leq l\leq k$. We know that \cite[p.165]{Lov:12} 
\begin{align} \label{eq:mfc}
\E[\abs*{\frac{|S_{m,n}\cap I_l|}{n}-\frac{1}{k}}] \leq \frac{1}{kn^{3/8}}.
\end{align}
We modify $S_{m,n}$ to construct a new $S_{m,n}' = \{x_{m,1}'',\ldots, x_{m,n}''\}$ by: (a) only shifting of points in $S_{m,n}$ without changing their ordering, (b) making $|S_{m,n}'\cap I_l| = \floor{n/k}$ or $\floor{n/k}+1$, and (c) minimizing the symmetric set difference $|S_{m,n}'\Delta S_{m,n}|$ (illustrated in \cref{fig:smn}). By \cref{eq:mfc}, we have
\begin{align*}
\E[|S_{m,n}'\Delta S_{m,n}|] \leq \frac{2n}{kn^{3/8}} + \frac{4n}{kn^{3/8}} + \cdots + \frac{2kn}{kn^{3/8}} \leq \frac{2kn}{n^{3/8}}. 
\end{align*}

\begin{figure}[!htb]
\centering
\includegraphics[scale=0.6]{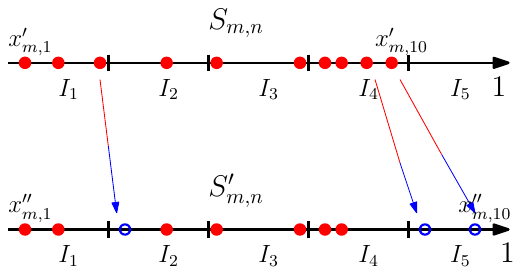}
\caption{Illustration of how $S_{m,n}'$ is obtained from $S_{m,n}$ by shifting points.} 
\label{fig:smn}
\end{figure}

We now construct $g_{m,n,k}$ 
(cf.\ (\ref{eq:fmg})) using $S_{m,n}'$: subdivide $[0,1]$ into $n$ equal intervals $J_1,\ldots, J_n$, and $g_{m,n,k}(x) = f_m'(x_{m,i}'')$ if $x\in J_i$. Notice that $g_{m,n,k}(x) = f_{m,n}(x)$ on $J_i$ if $x_{m,i}' = x_{m,i}''$. 
On average, $f_{m,n}$ and $g_{m,n,k}$ differs on at most $2kn/n^{3/8}$ subintervals $J_i$ (of size $1/n$). Hence, 
\begin{align*}
\E\norm{f_{m,n}-g_{m,n,k}}_1\leq 2kn/n^{3/8}\cdot 1/n\cdot 2b_f = 4kb_f/n^{3/8}.
\end{align*}
It converges to $0$ if we first fix $k$ and then let $n\to \infty$. It thus suffices to compare $g_{m,n,k}$ and $f_m'$, and intuitively the former is essentially a Riemann sum of the latter.

On each interval $I_l$, let $\sup_l$ and $\inf_l$ be $\sup_{x\in I_l}f_m'(x)$ and $\inf_{x\in I_l}f_m'(x)$ respectively. For each $x\in I_l$, we have 
\begin{align*}
|g_{m,n,k}(x) - f_m'(x)| \leq \sup_l-\inf_l. 
\end{align*}
Moreover, by (b) above, the number of points in $S_{m,n}'\cap I_l$ deviates from the average by at most $1$. Hence on $I_l$, we have
\begin{align*}
\int_{I_l}|g_{m,n,k}(x)-f_m'(x)|\ud x \leq \frac{1}{k}(\sup_l-\inf_l) + 2\cdot \frac{2b_f}{n}.
\end{align*}
The factor $2$ (in front of $2b_f/n$) is due to that $I_l$ may contain an extra point and have an additional contribution from either $I_{l-1}$ or $I_{l+1}$. Summing over all $I_l, 1\leq i\leq k$, we have
\begin{align*}
\norm{g_{m,n,k}-f_m'}_1 \leq \sum_{1\leq l\leq k}|I_l|(\sup_l-\inf_l) + \frac{4kb_f}{n}. 
\end{align*}
Therefore, as $f_m'$ is \gls{aects}, and hence Riemann integrable, if we first choose $k \to \infty$ to be sufficiently large and then let $n \to \infty$, the norm $\norm{g_{m,n,k}-f_m'}_1$ can be arbitrarily small. This proves the theorem. 
\end{IEEEproof}

\begin{IEEEproof}[Proof of \cref{coro:dwm}]
Consider $W_m$, $W_m'$ from \cref{eq:www}. For $m>0$, let the eigenvalues of the self-adjoint HS operator $T_{W_m}$ be 
\begin{align*}
\lambda_{-1,m} \leq \lambda_{-2,m} \leq \ldots \leq 0 \leq \ldots \leq \lambda_{2,m} \leq \lambda_{1,m}.
\end{align*}
Let $\overline{\lambda_t} = \lambda_t/\sqrt{\norm{W}_1}$ and $\overline{\lambda_{t,m}} = \lambda_{t,m}/\sqrt{\norm{W_m}_1}$, as eigenvalues of $T_{W^{\mathfrak{s}}}$ and $T_{W_m^{\mathfrak{s}}}$ respectively. We claim 
\begin{align}\label{eq:ltl}
\overline{\lambda_{t,m}} \to \overline{\lambda_t},\ \text{as } m \to \infty.
\end{align}
The Courant-Fischer min-max principle \cite[Ch.~12]{Lie01} says that, for any $t<0$, we have the eigenvalue $\overline{\lambda_t}$ of $T_{W^{\mathfrak{s}}}$ being 
\begin{align}\label{eq:lmp}
\min_{\phi_1,\ldots,\phi_{|t|}}\max \set*{\ip*{\phi}{T_{W^{\mathfrak{s}}}\phi} \given \phi\in B_1(\phi_1,\ldots,\phi_{|t|})}, 
\end{align}
where $B_1(\phi_1,\ldots,\phi_{|t|}) \subset \spn(\phi_1,\ldots,\phi_{|t|})$ consist of norm $1$ elements.

Notice only $t<0$ is considered as the min-max principle can only be applied to eigenvalues below the essential spectrum, which is $\{0\}$ for any compact operator.

For any $m$, and for any $\phi$ with $\norm{\phi}_2=1$ we have 
\begin{align*}
& |\langle \phi, T_{W^{\mathfrak{s}}}\phi \rangle - \langle \phi,T_{W_m^{\mathfrak{s}}}\phi \rangle| = |\langle \phi, (T_{W^{\mathfrak{s}}}-T_{W_m^{\mathfrak{s}}})\phi \rangle|
\\&\leq \norm{\phi}_2\norm{(T_{W^{\mathfrak{s}}}-T_{W_m^{\mathfrak{s}}})\phi}_2  \leq \norm{T_{W^{\mathfrak{s}}}-T_{W_m^{\mathfrak{s}}}}_{2,2}.
\end{align*}
We have seen in \cref{lem:fta} that $W_m^{\mathfrak{s}}$ converges to $W^{\mathfrak{s}}$ in $1$-norm as $m\to\infty$, which implies convergence in the HS and hence the operator norm $\norm{\cdot}_{2,2}$.  Therefore, for $m$ sufficiently large so that the operator norm of $T_{W^{\mathfrak{s}}}-T_{W_m^{\mathfrak{s}}}$ is sufficiently small, so is $|\langle \phi, T_{W^{\mathfrak{s}}}\phi \rangle - \langle \phi,T_{W_m^{\mathfrak{s}}}\phi \rangle|$ small. The claim holds for any $\phi$ with $\norm{\phi}_2=1$, hence $|\overline{\lambda_t}-\overline{\lambda_{t,m}}|$ can also be arbitrarily small by (\ref{eq:lmp}). Therefore, $\overline{\lambda_{t,m}} \to \overline{\lambda_t}$ as $m\to\infty$. For $t>0$, we can apply the same argument to $-T_{W^{\mathfrak{s}}}$. In summary, for $t\neq 0$, we have 
\begin{align*}
\frac{\lambda_{t,m}}{\sqrt{\norm{W_m}_1}} \to \frac{\lambda_t}{\sqrt{\norm{W}_1}},\ \text{as } m\to \infty.
\end{align*}

It remains to show that with probability one, for each $m>0$ and $t$, we have 
\begin{align} \label{eq:lnf}
\lim_{n\to \infty} \frac{\lambda_{t,m,n}}{\sqrt{2|E_{m,n}|}} = \frac{\lambda_{t,m}}{\sqrt{\norm{W_m}_1}}.
\end{align}
By \cite[Lemma 10.16]{Lov:12}, with probability one, the canonical graphons $(W_{m,n})_{n>0}$ of $(G_{m,n})_{n>0}$ converge to $W_m'$ in the cut distance. It follows that  $\sqrt{2|E_{m,n}|}/n \to \sqrt{\norm{W_m'}_1}$ as $n\to \infty$. Moreover, let $\lambda_{t,m}'$ be the $t$-th eigenvalue of $T_{W_m'}$. The spectral convergence result for classical graphons \cite[Lemma 4]{Lua21} or \cite[Theorem 6.7]{Bor12} implies $\lim_{n\to \infty} \lambda_{t,m,n}/n = \lambda'_{t,m}$. 

We also have that $W_m^{\mathfrak{s}} = W_m'^{\mathfrak{s}}$ by (\ref{eq:wmw}), and therefore, $T_{W_m^{\mathfrak{s}}}$ and $T_{W_m'^{\mathfrak{s}}}$ have the same spectrum. This implies that $\lambda_{t,m}/\sqrt{\norm{W_m}_1}$ is the same as $\lambda_{t,m}'/\sqrt{\norm{W_m'}_1}$. We conclude
\begin{align*}
    \lim_{n\to \infty} \frac{\lambda_{t,m,n}/n}{ \sqrt{2|E_{m,n}|}/n} = \frac{\lambda_{t,m}'}{\sqrt{\norm{W_m'}_1}} = \frac{\lambda_{t,m}}{\sqrt{\norm{W_m}_1}}.
\end{align*}  
\end{IEEEproof}

\begin{IEEEproof}[Proof of \cref{coro:ihi}]
Suppose $h$ is continuous on the closure $\overline{U}$ of $U$, a bounded open interval containing the eigenvalues of $T_{W^{\mathfrak{s}}}$. By the Stone-Weierstrass theorem \cite{Rud76}, $h$ is a uniform limit of polynomials on $\overline{U}$. Thus, it suffices to show the theorem for the case where $h$ is the uniform limit of a sequence of polynomials $(P_l(\cdot))_{l\geq1}$ on $U$.

We first notice that $P_l(0) \to h(0)=0$ as $l \to \infty$. By replacing each $P_l$ with $P_l-P_l(0)$ if necessary, we assume $P_l(0)=0$. We claim that $U$ contains the eigenvalues of $T_{W_{m,n}^{\mathfrak{s}}}$ for $m,n$ sufficiently large. Notice $U$ is an open interval containing $\overline{\lambda_{-1}}$ and $\overline{\lambda_1}$. By (\ref{eq:ltl}), for $m,n$ sufficiently large, $U$ contains the largest and the smallest eigenvalues, and hence all eigenvalues of $T_{W_{m,n}^{\mathfrak{s}}}$. This proves the claim. 

We next show that for $T$ being either $T_{W^{\mathfrak{s}}}$ or $T_{W_{m,n}^{\mathfrak{s}}}$, we have $\lim_{l\to \infty} P_l(T) = h(T)$ uniformly (independent of $T$). Let the non-eigenvalues of $T$ be $(\lambda_{t,T})_{t\in \mathbb{Z}\backslash\{0\}}$ and the eigenvectors be $(\varphi_{t,T})_{t\in \mathbb{Z}\backslash\{0\}}$. Since $P_l$ converges to $h$ uniformly on $U$, for any $\epsilon>0$, there is an $l_0$ (independent of $T$) such that $|P_l(x)-h(x)| \leq \epsilon$ if $x\in U$ and $l\geq l_0$. By the previous paragraph, each $\lambda_{t,T}$ is contained in $U$. Hence, for any signal $f \in L^2(\mathbb{R}_+)$, we have
\begin{align*}
& \norm{P_l(T)f - h(T)f}_2 \\
& = \norm{\sum_{t\neq 0}\big(P_l(\lambda_{t,T}) - h(\lambda_{t,T})\big)\langle \varphi_{t,T}, f\rangle\varphi_{t,T} }_2   \\
&\leq \epsilon \norm{\sum_{t\neq 0}\langle \varphi_{t,T}, f\rangle\varphi_{t,T}}_2 = \epsilon \norm{f}_2. 
\end{align*}
The claim follows. The uniform convergence is needed to apply the Moore-Osgood theorem \cite{Tay12b} for interchanging iterated limits, i.e., exchanging $\lim_{l \to \infty}$ with $(\lim_{m\to\infty}\lim_{n\to\infty})$.

On the other hand, for fixed $l>0$, by \cref{coro:lpb} and \cref{thm:swi}, we have $\lim_{m \to\infty}\lim_{n \to\infty} P_l(T_{W_{m,n}^{\mathfrak{s}}}) = P_l(T_{W^{\mathfrak{s}}})$ with probability one. 

We now combine everything to obtain the claimed (\ref{eq:lml})
\begin{align*}
&\lim_{m\to\infty}\lim_{n\to\infty} h(T_{W_{m,n}^{\mathfrak{s}}}) = \lim_{m\to\infty}\lim_{n\to\infty} \lim_{l\to\infty}P_l(T_{W_{m,n}^{\mathfrak{s}}}) \\
&= \lim_{l\to\infty}\lim_{m\to\infty}\lim_{n\to\infty} P_l(T_{W_{m,n}^{\mathfrak{s}}}) = \lim_{l\to \infty} P_l(T_{W^{\mathfrak{s}}}) = h(T_{W^{\mathfrak{s}}}), 
\end{align*}
where the $2$nd equality is due to the Moore-Osgood theorem. 

\end{IEEEproof}

\bibliographystyle{IEEEtran}
\bibliography{IEEEabrv,StringDefinitions,bib/refs,bib/SIGNAL}

\end{document}